\documentclass[review]{elsarticle}

\usepackage{geometry}
\usepackage{amsmath}
\usepackage{amssymb}
\usepackage{graphicx}

\usepackage{bm}
\usepackage{color}
\usepackage{multirow}
\usepackage{makecell} % cell newline

\journal{Signal Processing}

\begin{document}

\begin{frontmatter}

\title{Study of Frequency domain exponential functional link network filters}

\author[label1]{Tao Yu\corref{cor1}}
\cortext[cor1]{Corresponding author}
\ead{yutao@swpu.edu.cn}
\author[label1]{Shijie Tan}
\ead{tan\_edu@163.com}
\address[label1]{School of Electrical Engineering and Information, Southwest Petroleum University, Chengdu 610500, China}
\author[label2,label3]{Rodrigo C. de Lamare}
\ead{delamare@puc-rio.br}
\address[label2]{CETUC, Pontifical Catholic University of Rio de Janeiro, Rio de Janeiro 22451-900, Brazil}
\address[label3]{Department of Electronic Engineering, University of York, York YO10 5DD, U.K.}
\author[label4]{Yi Yu}
\ead{yuyi\_xyuan@163.com}
\address[label4]{School of Information Engineering, Robot Technology Used for Special Environment Key Laboratory of Sichuan Province, Southwest University of Science and Technology, Mianyang 621010, China}

\begin{abstract}
The exponential functional link network (EFLN) filter has attracted tremendous interest due to its enhanced nonlinear modeling capability. However, the computational complexity will dramatically increase with the dimension growth of the EFLN-based filter. To improve the computational efficiency, we propose a novel frequency domain exponential functional link network (FDEFLN) filter in this paper. The idea is to organize the samples in blocks of expanded input data, transform them from time domain to frequency domain, and thus execute the filtering and adaptation procedures in frequency domain with the overlap-save method. A FDEFLN-based nonlinear active noise control (NANC) system has also been developed to form the frequency domain exponential filtered-s least mean-square (FDEFsLMS) algorithm. Moreover, the stability, steady-state performance and computational complexity of algorithms are analyzed. Finally, several numerical experiments corroborate the proposed FDEFLN-based algorithms in nonlinear system identification, acoustic echo cancellation and NANC implementations, which demonstrate much better computational efficiency.
\end{abstract}


\end{frontmatter}

\section{Introduction}

Nonlinear filtering techniques have been extensively investigated in
the last decade, mainly due to their ability to deal with the
inherent nonlinearities of practical systems. Applications of
nonlinear adaptive filters, which possess the learning capability to
obtain nonlinear system parameters, have been related to system
identification \cite{Carini2019Nonlinear}, echo cancellation
\cite{Comminiello2014Nonlinear}, noise control \cite{Lu2021Survey},
and acoustic feedback cancellation in hearing aids
\cite{Bhattacharjee2021Fast}. For filtering-oriented purposes,
several adaptive learning algorithms
\cite{intadap,jio,jidf,jidfecho,jiols,mcg,jiomimo,sjidf,l1stap,smtvb,smce,saalt,damdc,locsme,lrcc,dce,okspme,aaidd,dlmme,dynovs,1bitce,dqalms,dlmm,rdlms,lbal}
have emerged within a wide range of linear and nonlinear systems.

Volterra adaptive filtering (VAF) has been early used to model nonlinear characteristics by the Volterra series expansion \cite{Tan2001Adaptive}. However, VAF with more free parameters often has high computational burden and analytical difficulty. As widely known, neural networks possess the universal approximation property and can achieve nonlinear mapping more flexibly, but its computational cost is very high and it is easy to get trapped into local solution \cite{Zhao2011Low}. By virtue of the kernel trick, kernel adaptive filtering (KAF) maps the inputs to a high-dimensional feature space \cite{Flores2019Set}. Nevertheless, the extra degrees of freedom of KAF also increase the computational load. Spline adaptive filtering (SAF) is structured as a cascade of a linear combiner and an adaptive local spline interpolation. Although SAF is flexible and simple to implement, it requires a priori knowledge of piecewise nonlinear regression \cite{Scarpiniti2013Nonlinear,Patel2016Compensating,Yu2021RobustSpline}.

Another popular linear-in-parameters (LIP) nonlinear filtering scheme is referred to as the functional link network (FLN), which comprises a nonlinear functional expansion block followed by a finite impulse response (FIR) filter \cite{Sicuranza2012BIBO}. In the FLN-based filter, the input data is expanded to a higher-dimensional nonlinearity including the trigonometric \cite{Das2004Active}, Hermite \cite{Yin2020Hermite}, and Chebyshev \cite{Carini2016Study} nonlinear series. The most conventional filtering scheme among them is the trigonometric functional link network (TFLN) consisting of pure trigonometric basis functions, which has mild computational requirement and efficient modeling capability \cite{Sicuranza2012BIBO}. As proved in \cite{Hermus2005Perceptual}, the exponentially varying trigonometric series can further improve the modeling ability of nonlinear systems. Therefore, an exponential functional link network (EFLN) filter was designed in \cite{Patel2016Design}, where the input signal was expanded by the exponentially varying trigonometric polynomials, and the least mean-square (LMS) approach was utilized to learn the FIR coefficients and the exponential factor of EFLN. The convergence behavior and performance analysis of the EFLN-based filter have been studied in \cite{Patel2021Convergence}. Following this study, the EFLN algorithm was extended to promote the convergence property and modeling accuracy, which further demonstrated that EFLN-based filters hold advantages in nonlinear filtering tasks \cite{Zhang2018Recursive,Deb2020Design,Bhattacharjee2020Nonlinear}. Recently, in order to perform well in the presence of impulsive noise, a robust EFLN algorithm exploiting the inverse square root cost function was proposed, and its convergence analysis was reported in \cite{Yu2021RobustAdaptive}.

Common feed-forward nonlinear active noise control (NANC) systems for suppressing acoustic noises consist of a reference microphone and an error microphone, and an active loudspeaker driven by the control mechanism to generate an antinoise signal \cite{George2013Advances}. In practical NANC systems, the captured reference noise may be a nonlinear noise sequence, and the primary path and secondary path often suffer from nonlinear distortions. Consequently, nonlinear adaptive filtering techniques can provide powerful active control mechanisms for suppressing noises in NANC systems. As a well-known way of compensating nonlinear distortions, the filtered-s least mean-square (FsLMS) algorithm based on the traditional TFLN filter has been developed in \cite{Das2004Active}. This adaptive control scheme and its improved versions have drawn increasing attention and include various implementations in the field of NANC \cite{Zhou2007Efficient,George2012Robust,Luo2018Improved}. To enhance the modeling accuracy and noise mitigation capability, the work in \cite{Patel2016Design} also proposed the exponential filtered-s least mean-square (EFsLMS) algorithm based on EFLN, and its bound on learning rates has been derived. Later on, a generalized EFLN filter exploiting cross-terms with the channel-reduced diagonal structure was developed in NANC systems \cite{Le2018Generalized}. A review of NANC methodologies has been summarized in \cite{George2013Advances,Lu2021Survey} and references therein.

However, it is noteworthy that the nonlinear functional expansion often results in expanded inputs with higher-dimensional nonlinearities, and the number of filter weights may become excessive large. In particular, with the dimension growth of the filter, the computational burden of the algorithm will go up dramatically. Time domain convolution and correlation operations can be realized by the fast Fourier transform (FFT) \cite{Haykin2014Adaptive}, thereby decreasing the complexity of adaptive algorithms. Hence, several frequency domain filtering algorithms were developed, and a unified framework was constructed to analyze the convergence behaviors of linear adaptive filters in frequency domain \cite{Yang2019Mean,Yang2019Unified}. Due to their computational advantage, frequency domain adaptive filtering has been applied in linear noise suppressing systems \cite{Das2007New,Yang2020Stochastic,Shi2021Comb}. Recent work in \cite{Yang2020Frequency} proposed the frequency domain spline adaptive filtering with reduction of computational demand for nonlinear system identification (NSI). In realistic acoustic scenarios, there commonly exist nonlinear distortions of the loudspeaker or microphone module. Therefore, nonlinear filtering algorithms in nonlinear acoustic echo cancellation (NAEC) applications have attracted a growing attention. In the early stages, a nonlinear power filtering scheme has been considered for NAEC, whose echo path has been modeled by a Hammerstein structure of the cascade of a nonlinear polynomial followed by a linear filter \cite{Kuech2005Nonlinear}, which also mentioned power filtering as a special case of Volterra filtering in diagonal representations. On this basis, a frequency domain power filtering (FDPF) algorithm was developed by combining the discrete Fourier transform (DFT) implementation of power filters with the overlap-save method \cite{Kuech2006Orthogonalized}. The block implementation of power filters for the DFT approach can be considered as the inherent multi-channel structure, and thus the adaptation can be performed independently for each channel \cite{Kuech2007Nonlinear}. In order to improve the computational efficiency for acoustic echo cancellers, the Kalman filter based on the state-space model of the echo path was formulated entirely in frequency domain \cite{Enzner2006Frequency}. Additionally, the work in \cite{Malik2011Fourier} established a class of FDPF structures to model Hammerstein nonlinearities for acoustic system identification. Later on, by absorbing the coefficients of nonlinear expansion into the echo path, the cascade model is transformed into an equivalent multi-channel structure, and the frequency domain Kalman filtering algorithm has been proposed in \cite{Malik2012State}. In NAEC environments for residual echo suppression, state-space frequency domain adaptive algorithms by combining Kalman filter and nonlinear expansion to exploit the advantages of reducing computational costs, in which the formulations incorporating memory or memoryless nonlinearities have been implemented in DFT domain \cite{Malik2013Variational,Vogt2019State}. The majority of the above-mentioned LIP nonlinear filters are adjusted by utilizing time domain adaptive filtering, whereas they may require extensive computational resources. There are numerous developments of LIP nonlinear filters in frequency domain, where the functional expansion block was followed by block linear filter using DFT implementation with the overlap-save method. Some of the most popular families of LIP nonlinear filters through DFT are based on the Volterra expansion and the trigonometric expansion \cite{Kuech2005Partitioned,Zeller2010Fast,Park2014Frequency,Das2006Development}. Although the importance of computational efficiency for LIP nonlinear filters has been recognized, work on the frequency domain filtering scheme has been limited so far, mainly due to some challenges, such as the frequency domain adaptation after the functional transformation.

In this paper, we develop a class of frequency domain filtering schemes based on EFLN, and apply it to nonlinear filtering tasks. The main contributions are:
\begin{enumerate}[1)]
  \item We propose a novel frequency domain exponential functional link network (FDEFLN) filtering scheme to achieve better computational efficiency than EFLN. Moreover, a FDEFLN-based NANC system has also been developed to form the frequency domain exponential filtered-s least mean-square (FDEFsLMS) algorithm.
  \item The stability, steady-state performance and computational complexity of the proposed frequency domain algorithms are derived and analyzed. The convergence conditions for both the weights and the exponential factor are given according to the energy conservation principle. In addition, the theoretical analysis of the adaptive algorithm is carried out in the mean-square sense, and a closed-form expression of the steady-state performance is given.
  \item Numerical experiments verify the superiority of the proposed algorithms in NSI, NAEC and NANC implementations to demonstrate the significant advantages of reducing computational costs on the premise of good convergence performance.
\end{enumerate}

This paper is arranged as follows. Some preliminaries of EFLN are introduced in Section 2. Section 3 presents the proposed FDEFLN nonlinear filtering scheme in detail. The FDEFLN-based filtering scheme is applied to NANC in Section 4 and the FDEFsLMS algorithm is also developed. The stability, steady-state performance and computational complexity are given in Section 5. Numerical experiments are carried out in Section 6. This paper ends with the conclusion in Section 7.

\section{Preliminaries}

\begin{figure}[!ht]
\centering
\includegraphics[width=2.5in]{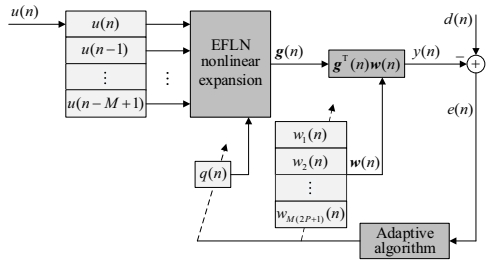}
\caption{Schematic of the EFLN-based filter.}
\label{fig:EFLN_model}
\end{figure}
The schematic of the EFLN-based filter is illustrated in Fig. \ref{fig:EFLN_model}, whose scheme comprises an EFLN nonlinear expansion block followed by a linear FIR filter \cite{Patel2016Design}. An $M$-dimensional tapped delay input vector is defined as $\bm u(n)=[u(n),u(n-1),\dots,u(n-M+1)]^\mathrm{T} \in \mathbb{R}^M$ with $u(n)$ being the input signal at time $n$. Considering a $P$-order functional expansion of EFLN, the $M$-dimensional input vector $\bm u(n)$ is expanded to an $M(2P+1)$-dimensional expanded input vector $\bm g(n)$ as
\begin{align}
\bm g(n) = \big[\bm g_1^\mathrm{T}(n),\bm g_2^\mathrm{T}(n),\dots, \bm g_{2P+1}^\mathrm{T}(n)\big]^\mathrm{T} \in \mathbb{R}^{M(2P+1)}
\end{align}
where its subvectors $\bm g_i(n) = [g_i(n),g_i(n-1),\dots,g_i(n-M+1)]^\mathrm{T} \in \mathbb{R}^M, i = 1,2,\dots,2P+1$ with the exponential factor $q(n)$ are formulated as
\begin{align}
\bm g_i(n) =
\left\{\begin{aligned}
&\bm u(n),  & &  i = 1\\
&e^{-q(n)|\bm u(n)|} \circ \sin[p \pi \bm u(n)], & &  i = 2p\\
&e^{-q(n)|\bm u(n)|} \circ \cos[p \pi \bm u(n)], & &  i = 2p+1
\end{aligned}\right.
\end{align}
where $p = 1,2,\dots,P$ is the expansion index, and the symbol of $\circ$ represents the Hadamard product.

The filtered output signal $y(n)$ is given by
\begin{align}
y(n) = \bm g ^\mathrm{T}(n)\bm w(n)
\end{align}
where the weight vector with the length $M(2P+1)$ is $\bm w(n)=[w_1(n),w_2(n),\dots,w_{M(2P+1)}(n)]^\mathrm{T} \in \mathbb{R}^{M(2P+1)}$. In the process of identification, the noisy desired signal is expressed as
\begin{align}
d(n) = \bar y(n) + \eta(n)= \bar {\bm g}^\mathrm{T}(n) \bar{\bm w} + \eta(n)
\end{align}
where $\bar y(n) = \bar {\bm g}^\mathrm{T}(n)\bar{\bm w}$ denotes the system output signal, $\bar {\bm g}(n)$ is the expanded input vector with the ideal exponential factor $\bar q$, $\bar{\bm w}$ is the ideal weight vector, and $\eta(n)$ is a zero-mean additive noise. The error signal $e(n)$ is defined by
\begin{align}
e(n) = d(n) - y(n) = d(n) - \bm g ^\mathrm{T}(n)\bm w(n)
\end{align}
where $d(n)$ is the noisy desired signal.

The EFLN filtering scheme always solves the mean-square error (MSE) minimization problem to get the corresponding adaptive solution, thereby updating the weights and the exponential factor adjusting the magnitude of trigonometric series. Taking advantage of the stochastic gradient descent (SGD) technique, the adaptation rules of the weight vector and exponential factor can be adjusted as
\begin{align}\label{eq:EFLN_update}
\bm w(n+1) &= \bm w (n) + \mu_w e(n) \bm g(n) \nonumber \\
q(n+1) &= q(n) + \mu_q e(n) \bm h^\mathrm{T} (n) \bm w(n)
\end{align}
where $\mu_w$ and $\mu_q$ represent the step sizes, and the vector $\bm h (n)$ is calculated as
\begin{align}
\bm h(n) &= \frac{ \partial \bm g(n) }{ \partial q(n) } = \big[\bm h_1^\mathrm{T}(n),\bm h_2^\mathrm{T}(n),\dots, \bm h_{2P+1}^\mathrm{T}(n)\big]^\mathrm{T} \in \mathbb{R}^{M(2P+1)}
\end{align}
with its elements $\bm h_i(n) =\frac{\partial \bm g_i(n)}{\partial q(n)} = [h_i(n),h_i(n-1),\dots,h_i(n-M+1)]^\mathrm{T}\in \mathbb{R}^M$ being formulated as
\begin{align}
\bm h_i(n) =
\left\{\begin{aligned}
&\bm 0,  & &  i = 1\\
&-|\bm u(n)| \circ e^{-q(n)|\bm u(n)|} \circ \sin[p \pi \bm u(n)], & &  i = 2p\\
&-|\bm u(n)| \circ e^{-q(n)|\bm u(n)|} \circ \cos[p \pi \bm u(n)], & &  i = 2p+1
\end{aligned}\right.
\end{align}
where $\bm 0$ represents a zero vector or a zero matrix with appropriate dimension.

Obviously, the EFLN algorithm performs the filtering and adaptation processes at every time $n$, which updates from the current weights and exponential factor to yield their successive processes. Therefore, the computational complexity for each iteration will dramatically increase with the long length of the FIR filter. We thus propose a novel frequency domain implementation which possesses a reduced computational demand as compared to its time domain counterpart.

\section{Proposed FDEFLN nonlinear filtering scheme}

In this section, we introduce the proposed FDEFLN nonlinear filtering scheme, which can be realized with the block EFLN-based scheme via the overlap-save method.

\begin{figure}[!ht]
\centering
\includegraphics[width=4.4in]{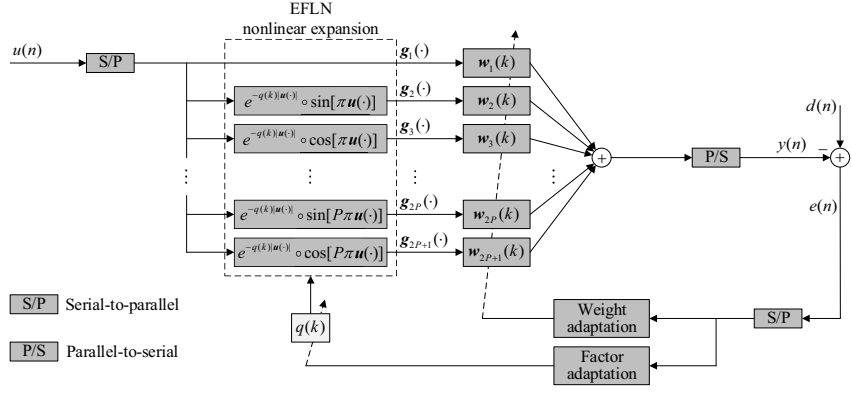}
\caption{Schematic of the block EFLN-based filter.}
\label{fig:BEFLN_model}
\end{figure}
To overcome the computational burden, a block nonlinear filtering implementation based on EFLN is initially depicted in Fig. \ref{fig:BEFLN_model}, which consists of the block filters with the inputs modulated by the exponentially varying trigonometric basis functions. In this block EFLN filtering scheme, the input data $u(n)$ is split into $M$-point blocks relying on a serial-to-parallel converter, and after the nonlinear expansion of each $M$-sample block, adaptations of the filter proceed on the block-by-block way instead of the sample-by-sample way as in the EFLN filter.

Let the block index $k$ relate to the initial sample time $n$ as $n = kM+j, j = 1,2,\dots,M$, in which $M$ is the block length and is equal to the tapped delay length. The expanded input data for block $k$ can be written in a matrix form as follows
\begin{align}
\bm G(k) &= [\bm g(kM+1), \bm g(kM+2), \dots, \bm g(kM+M)] \nonumber \\
&= \left[ \begin{array}{*{20}{c}}
\bm g_1(kM+1)        &\dots   &\bm g_1(kM+M)       \\
\bm g_2(kM+1)        &\dots   &\bm g_2(kM+M)       \\
\vdots               &\ddots  &\vdots              \\
\bm g_{2P+1}(kM+1)   &\dots   &\bm g_{2P+1}(kM+M)
\end{array} \right] = \left[ \begin{array}{*{20}{c}}
\bm G_1(k)        \\
\bm G_2(k)        \\
\vdots               \\
\bm G_{2P+1}(k)
\end{array} \right] \in \mathbb{R}^{M(2P+1) \times M}
\end{align}
where $\bm G_i(k) = [\bm g_i(kM+1),\bm g_i(kM+2),\dots,\bm g_i(kM+M)] \in \mathbb{R}^{M \times M}$ for each $i=1,2,\dots,2P+1$. Over this block of expanded input data, the weight vector of the filter is held at the value $\bm w(k)$, which is a rewrite of $\bm w(n)$ for $n = k$ as
\begin{align}
\bm w(k)&=[w_1(k),w_2(k),\dots,w_{M(2P+1)}(k)]^\mathrm{T} \nonumber \\
&= [\bm w_1^\mathrm{T}(k),\bm w_2^\mathrm{T}(k),\dots,\bm w_{2P+1}^\mathrm{T}(k)]^\mathrm{T} \in \mathbb{R}^{M(2P+1)}
\end{align}
where $\bm w_i(k) = [w_{iM-M+1}(k),\dots,w_{iM}(k)]^\mathrm{T} \in \mathbb{R}^M$.

In order to implement the block EFLN filtering scheme in a computationally efficient manner, the frequency domain method through FFT strategy is further adopted, rather than performing the adaptations in time domain as described previously. It means that this idea is to organize the samples in blocks of expanded input data, thereby executing the filtering and adaptation procedures in frequency domain. It has been proved that the overlap-save method with 50\% overlap is the most efficient procedure for fast convolution, i.e., the block length is equal to the tapped delay length \cite{Haykin2014Adaptive}. Henceforth, we concentrate on the 50\% overlap-save method to implement the FDEFLN filtering scheme.

At block $k$, the input sequence $u(n)$ is cached in the data block $\bm u(k) = [u(kM+1),u(kM+2),\dots,u(kM+M)]^\mathrm{T} \in \mathbb{R}^M$ for every $M$ samples. For each $i=1,2,\dots,2P+1$, the expanded input signal $g_i(kM+j)$ will be cached in the following data block
\begin{align}
\bm g_i(k) = [g_i(kM+1),g_i(kM+2),\dots,g_i(kM+M)]^\mathrm{T} \in \mathbb{R}^M  .
\end{align}
We thus have
\begin{align}
\bm g_i(k) =
\left\{\begin{aligned}
&\bm u(k),  & &  i = 1\\
&e^{-q(k)|\bm u(k)|} \circ \sin[p \pi \bm u(k)], & &  i = 2p\\
&e^{-q(k)|\bm u(k)|} \circ \cos[p \pi \bm u(k)], & &  i = 2p+1   .
\end{aligned}\right.
\end{align}

The filtered output in response to the expanded input signal is defined by
\begin{align}
y(kM+j) =&~ \bm g ^\mathrm{T}(kM+j)\bm w(k)  \nonumber\\
=&~ \bm g_1 ^\mathrm{T}(kM+j)\bm w_1(k)  + \cdots + \bm g_{2P+1} ^\mathrm{T}(kM+j)\bm w_{2P+1}(k) \nonumber\\
=&~\sum_{i=1}^{2P+1} y_i(kM+j)
\end{align}
where the exponential factor is held at the value $q(k)$ over the block $k$, and we denote
\begin{align}\label{eq:conv}
y_i(kM+j) &= \bm g_i ^\mathrm{T}(kM+j)\bm w_i(k) \nonumber\\
&=\sum_{\ell=1}^M g_i(kM+j-\ell+1) w_{iM-M+\ell}(k)   .
\end{align}

Taking the overlap-save method and the FFT of two successive blocks of data buffer yields a $2M$-dimensional vector
\begin{align}
\bm g_i(\tilde k) = \mathrm{FFT} \left[ \begin{array}{*{20}{c}}
\bm g_i(k-1)  \\
\bm g_i(k)
\end{array} \right]  = \mathrm{FFT} \left[ \begin{array}{*{20}{c}}
g_i(kM-M+1)  \\
\vdots \\
g_i(kM+M)
\end{array} \right] \in \mathbb{C}^{2M}
\end{align}
where $\mathrm{FFT}[\cdot]$ denotes the FFT operation, and $\tilde k$ represents the index in frequency domain. The weight vector $\bm w_i(k)$ is padded
with the equal number of zeros, and a zero-added weight vector of $2M$-points FFT coefficients is used as
\begin{align}
\bm w_i(\tilde k) = \mathrm{FFT} \left[ \begin{array}{*{20}{c}}
\bm w_i(k)  \\
\bm 0
\end{array} \right] \in \mathbb{C}^{2M}  .
\end{align}
Therefore, applying the overlap-save method to the convolution of \eqref{eq:conv} establishes an $M$-dimensional filtered output vector
\begin{align}
\bm y_i(k) &= [y_i(kM+1),y_i(kM+2),\dots,y_i(kM+M)]^\mathrm{T} \nonumber\\
&=\mathrm{IFFT}[\bm g_i(\tilde k) \circ \bm w_i(\tilde k) ]~\text{last}~M~\text{elements}
\end{align}
where $\mathrm{IFFT}[\cdot]$ denotes the inverse FFT operation. Only the last $M$ elements are stored, because the first $M$ elements are relevant to a circular convolution and should be discarded. Then, the filtered output vector for each data block $k$ can be given by
\begin{align}
\bm y(k) &= \bm G^\mathrm{T}(k) \bm w(k) \nonumber\\
&= [y(kM+1),y(kM+2),\dots,y(kM+M)]^\mathrm{T} \nonumber\\
&=\bm y_1(k) + \bm y_2(k) + \cdots + \bm y_{2P+1}(k)  \in\mathbb{R}^M  .
\end{align}

We will derive the adaptive learning rules for the weights and exponential factor by using the frequency domain method in the following. At block $k$, denote the desired signal vector as $\bm d(k) = [d(kM+1),d(kM+2),\dots,d(kM+M)]^\mathrm{T} \in\mathbb{R}^M$. The corresponding error signal vector is written as
\begin{align}
\bm e(k) &= [e(kM+1),e(kM+2),\dots,e(kM+M)]^\mathrm{T} \nonumber\\
& = \bm d(k) - \bm y(k)  \in\mathbb{R}^M  .
\end{align}

Also based on the MSE criterion, the cost function $J(k)$ is taken as
\begin{align}
J(k) = \|\bm e(k)\|^2 = \sum_{j=1}^M e^2(kM+j)
\end{align}
where $\|\cdot\|$ denotes the $l_2$-norm of its vector argument. The estimate of the gradient vector with respect to $\bm w_i(k)$ can be deduced as
\begin{align}\label{eq_wdot_freq}
\frac{\partial J(k)}{\partial \bm w_i(k)} & = -2 \sum_{j=1}^M e(kM+j) \bm g_i (kM+j) \nonumber\\
& = -2 \bm G_i(k) \bm e(k)   .
\end{align}
Through the zero-added error vector of $2M$-points FFT operation, the error signal vector is implemented in frequency domain as
\begin{align}
\bm e(\tilde k) = \mathrm{FFT} \left[ \begin{array}{*{20}{c}}
\bm 0  \\
\bm e(k)
\end{array} \right] \in \mathbb{C}^{2M}  .
\end{align}
Applying the overlap-save method to the correlation of \eqref{eq_wdot_freq} establishes an $M$-dimensional gradient vector
\begin{align}
\bm \phi_i(k) =\mathrm{IFFT}[\bm e(\tilde k) \circ \mathrm{conj}(\bm g_i(\tilde k) )]~\text{first}~M~\text{elements}
\end{align}
where $\mathrm{conj}(\cdot)$ denotes the complex conjugation operation, and obviously the first $M$ elements should be retained. By taking the SGD method, the learning rule of the $2M$-dimensional weight vector in frequency domain is achieved as
\begin{align}\label{eq:wrule}
\bm w_i(\tilde k +1) = \bm w_i(\tilde k) + \mu_w \mathrm{FFT} \left[ \begin{array}{*{20}{c}}
\bm \phi_i(k)  \\
\bm 0
\end{array} \right]  .
\end{align}

Additionally, the estimate of the exponential factor can be obtained as
\begin{align}
\frac{\partial J(k)}{\partial q(k)} & = -2 \sum_{j=1}^M e(kM+j) \sum_{i=1}^{2P+1} \bm h_i ^\mathrm{T}(kM+j) \bm w_i(k) \nonumber\\
& = -2 \sum_{j=1}^M e(kM+j) \sum_{i=1}^{2P+1} z_i (kM+j) \nonumber\\
& = -2 \sum_{j=1}^M e(kM+j) z (kM+j) \nonumber\\
& = -2\bm z^\mathrm{T}(k) \bm e(k)
\end{align}
where we denote
\begin{align}
\bm z(k) &= [z(kM+1),z(kM+2),\dots,z(kM+M)]^\mathrm{T} \nonumber\\
&=\bm z_1(k) + \bm z_2(k) + \cdots + \bm z_{2P+1}(k) \in\mathbb{R}^M
\end{align}
with $\bm z_i(k) = [z_i(kM+1),z_i(kM+2),\dots,z_i(kM+M)]^\mathrm{T}$ for each $i = 1,2,\dots,2P+1$ and $z (kM+j) = \sum_{i=1}^{2P+1} z_i (kM+j)$ for each $j=1,2,\dots,M$, and we have
\begin{align}\label{eq:z_conv}
z_i(kM+j) &= \bm h_i ^\mathrm{T}(kM+j)\bm w_i(k) \nonumber \\
&=\sum_{\ell=1}^M h_i(kM+j-\ell+1) w_{iM-M+\ell}(k)  .
\end{align}
Similar to the procedures in frequency domain, and taking the overlap-save method to the convolution of \eqref{eq:z_conv}, we can obtain
\begin{align}
\bm z_i(k) = \mathrm{IFFT}[\bm h_i(\tilde k) \circ \bm w_i(\tilde k) ]~\text{last}~M~\text{elements}
\end{align}
where we compute
\begin{align}
\bm h_i(\tilde k) &= \mathrm{FFT} \left[ \begin{array}{*{20}{c}}
\bm h_i(k-1)  \\
\bm h_i(k)
\end{array} \right]  \in \mathbb{C}^{2M}
\end{align}
with $\bm h_i(k) = [h_i(kM+1),h_i(kM+2),\dots,h_i(kM+M)]^\mathrm{T} \in \mathbb{R}^M$ for each data block $k$. Hence, the learning rule of the exponential factor can be updated as
\begin{align}\label{eq:qrule}
q(k+1) = q(k) + \mu_q \bm z^\mathrm{T}(k) \bm e(k)
\end{align}
by taking the SGD method.

\begin{figure*}[!ht]
\centering
\includegraphics[scale=0.46]{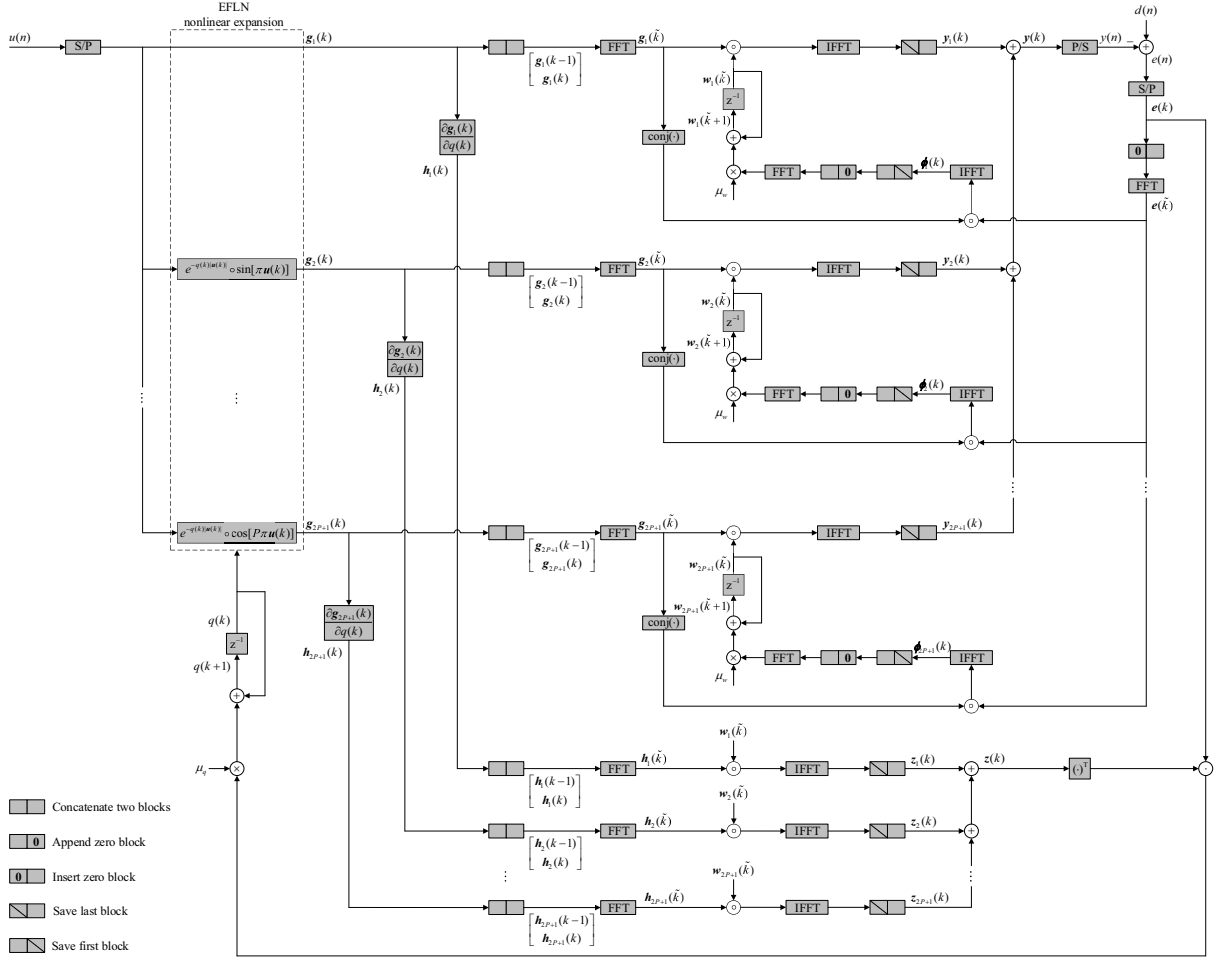}
\caption{Signal-flow graph of the FDEFLN algorithm.}
\label{fig:FDEFLN_model}
\end{figure*}
In the proposed FDEFLN algorithm for nonlinear filtering, after the EFLN nonlinear expansion of each data block of samples, the learning processes are performed in a block-by-block manner, and not by sample-by-sample fashion. Besides, on the basis of the 50\% overlap-save method, the samples in block of expanded input data are transformed to frequency domain ways, and the filtering and adaptation procedures are also performed in frequency domain. As a consequence, the computational demand of FDEFLN can be significantly reduced on the premise of convergence properties as compared to EFLN. Fig. \ref{fig:FDEFLN_model} illustrates a signal-flow graph of the FDEFLN filtering scheme with a frequency domain realization. To present the adaptive algorithm more clearly, the proposed FDEFLN algorithm is summarized in Table \ref{tab:FDEFLN}.
\begin{table}[!ht]\scriptsize
\centering
\renewcommand{\arraystretch}{1.2}
\setlength{\abovecaptionskip}{2pt}
\caption{Summary of the FDEFLN algorithm.}
\label{tab:FDEFLN}
\begin{tabular}{l}
\hline
\textbf{Initialization:} $\bm w(0), q(0), \mu_w, \mu_q$\\
\hline
  1:\hspace{6pt}\textbf{for} $k = 1,2,\dots$ \textbf{do}\\
  2:\hspace{12pt}\textbf{for} $i = 1,2,\dots,2P+1$ \textbf{do}\\
  3:\hspace{18pt}$\bm g_i(k) = [g_i(kM+1),g_i(kM+2),\dots,g_i(kM+M)]^\mathrm{T}$\\
  4:\hspace{18pt}$\bm g_i(\tilde k) = \mathrm{FFT} \left[ \begin{array}{*{20}{c}}
\bm g_i(k-1)  \\
\bm g_i(k)
\end{array} \right]$\\
  5:\hspace{18pt}$\bm y_i(k) =\mathrm{IFFT}[\bm g_i(\tilde k) \circ \bm w_i(\tilde k) ]$ last $M$ elements \\
  6:\hspace{12pt}\textbf{end for} \\
  7:\hspace{12pt}$\bm y(k) =\bm y_1(k) + \bm y_2(k) + \cdots + \bm y_{2P+1}(k)$\\
  8:\hspace{12pt}$\bm e(k) = \bm d(k) -\bm y(k)$\\
  9:\hspace{12pt}$\bm e(\tilde k) = \mathrm{FFT} \left[ \begin{array}{*{20}{c}}
\bm 0  \\
\bm e(k)
\end{array} \right]$\\
  10:\hspace{8pt}\textbf{for} $i = 1,2,\dots,2P+1$ \textbf{do}\\
  11:\hspace{14pt}$ \bm \phi_i(k) =\mathrm{IFFT}[\bm e(\tilde k) \circ \mathrm{conj}(\bm g_i(\tilde k) )]$ first $M$ elements \\
  12:\hspace{14pt}$\bm w_i(\tilde k +1) = \bm w_i(\tilde k) + \mu_w \mathrm{FFT} \left[ \begin{array}{*{20}{c}}
\bm \phi_i(k)  \\
\bm 0
\end{array} \right]$ \\
  13:\hspace{8pt}\textbf{end for} \\
  14:\hspace{8pt}\textbf{for} $i = 1,2,\dots,2P+1$ \textbf{do}\\
  15:\hspace{14pt}$\bm h_i(k) = [h_i(kM+1),h_i(kM+2),\dots,h_i(kM+M)]^\mathrm{T}$\\
  16:\hspace{14pt}$\bm h_i(\tilde k) = \mathrm{FFT} \left[ \begin{array}{*{20}{c}}
\bm h_i(k-1)  \\
\bm h_i(k)
\end{array} \right]$\\
  17:\hspace{14pt}$\bm z_i(k) =\mathrm{IFFT}[\bm h_i(\tilde k) \circ \bm w_i(\tilde k) ]$ last $M$ elements \\
  18:\hspace{8pt}\textbf{end for} \\
  19:\hspace{8pt}$\bm z(k) =\bm z_1(k) + \bm z_2(k) + \cdots + \bm z_{2P+1}(k)$\\
  20:\hspace{8pt}$q(k+1) = q(k) + \mu_q \bm z^\mathrm{T}(k) \bm e(k)$\\
  21:\hspace{2pt}\textbf{end for} \\
\hline
\end{tabular}
\end{table}

\section{FDEFLN-based NANC system}

\begin{figure}[!ht]
\centering
\includegraphics[width=4.5in]{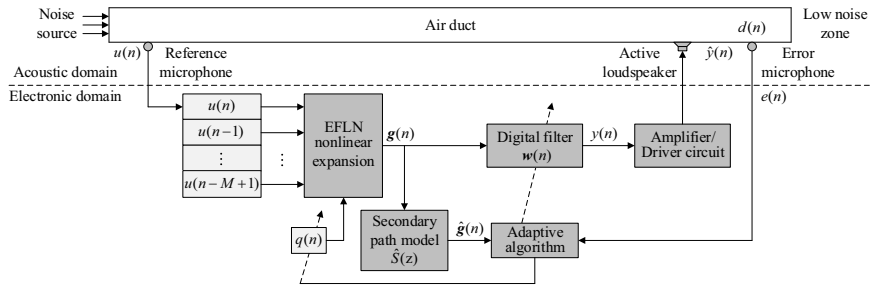}
\caption{Schematic of the EFLN-based NANC system.}
\label{fig:ANC_system}
\end{figure}
The schematic of the EFLN-based filter for the typical feed-forward NANC system is depicted in Fig. \ref{fig:ANC_system}. The reference signal $u(n)$ is acquired by a reference microphone near the noise source. $P(\mathrm{z})$ is the primary path model from the reference microphone to the error microphone. $S(\mathrm{z})$ with length $N$ is the secondary path model from the filtered output to the error microphone, and its impulse response is denoted as $s(n)$. $\hat S(\mathrm{z})$ represents the estimate of the secondary path model, which is generally assumed to be equal to $S(\mathrm{z})$ through a perfectly offline modeling \cite{George2013Advances,Patel2016Design}. $d(n)$ is the output of the primary path. $\hat y(n)=y(n)*s(n)$ is $y(n)$ filtered through $S(\mathrm{z})$, where the symbol of $*$ denotes the convolution operation. Henceforth, the superscript of \^{} represents the filtered version through the secondary path model. The residual noise $e(n)$ acquired by an error microphone is given by
\begin{align}
e(n) = d(n) - \hat y(n) = d(n) - y(n)*s(n)  .
\end{align}

In the EFLN-based NANC system, the weight vector $\bm w(n)$ and the exponential factor $q(n)$ are updated using the SGD technique, which minimize the MSE cost function to obtain
\begin{align}
\bm w(n+1) &= \bm w (n) + \mu_w e(n) \hat {\bm g}(n) \nonumber\\
q(n+1) &= q(n) + \mu_q e(n) \hat {\bm h}^\mathrm{T} (n) \bm w(n)
\end{align}
with $\hat{\bm g}(n) = \bm g(n)*s(n)$ and $\hat{\bm h}(n) = \bm h(n)*s(n)$ being the filtered versions through $\hat S(\mathrm{z})$, which form the EFsLMS algorithm.

Similar to the previous discussion, let the block index $k$ relate to the initial sample time $n$ as $n = kM+j, j = 1,2,\dots,M$ with $M$ being the block length, and the impulse response of the secondary path model is held at $s(k)$ over the block $k$. We then denote the filtered signal
\begin{align}\label{eq:conv2}
\hat g_i(kM+j) &= g_i(kM+j)*s(k)\nonumber\\
&=\sum_{\ell=1}^N g_i(kM+j-\ell+1) s_\ell(k)
\end{align}
where the elements of $\bm s(k) = [s_1(k),s_2(k),\dots,s_N(k)]^\mathrm{T} \in \mathbb{R}^N$ are the impulse response coefficients of the secondary path model. Hence, the filtered version of the expanded input data for block $k$ is written in matrix form as
\begin{align}
\hat{\bm G}(k) &= [\hat{\bm g}(kM+1), \hat{\bm g}(kM+2), \dots, \hat{\bm g}(kM+M)] = \left[ \begin{array}{*{20}{c}}
\bm G_1(k)*s(k)        \\
\bm G_2(k)*s(k)        \\
\vdots               \\
\bm G_{2P+1}(k)*s(k)
\end{array} \right]
\end{align}
where $\hat {\bm G}_i(k) = \bm G_i(k)*s(k) = [\hat{\bm g}_i(kM+1),\hat{\bm g}_i(kM+2),\dots,\hat{\bm g}_i(kM+M)]$ with $\hat{\bm g}_i(kM+j)=[\hat g_i(kM+j),\dots,\hat g_i(kM+j-M+1)]$. We define
\begin{align}
\hat y(kM+j) = y(kM+j)*s(k)= \sum_{i=1}^{2P+1} \hat y_i(kM+j)
\end{align}
with $\hat y_i(kM+j) = y_i(kM+j)*s(k)$. Let $d(kM+j)$ be the output signal, and the error signal $e(kM+j)$ is given by
\begin{align}
e(kM+j) &= d(kM+j) - \hat y(kM+j) \nonumber\\
&= d(kM+j) - \sum_{i=1}^{2P+1} \hat y_i(kM+j)  .
\end{align}

We then develop the FDEFsLMS algorithm based on FDEFLN for the NANC system. This frequency domain implementation is utilized to effectively calculate the relevant convolution and correlation operations in terms of block-by-block ways.

First applying the 50\% overlap-save method, the filtering of the expanded inputs through the estimate of secondary path is actualized in frequency domain for computational advantage. The zero-added $\bm s(k)$ of $2M$-points FFT coefficients is provided by
\begin{align}
\bm s(\tilde k) = \mathrm{FFT} \left[ \begin{array}{*{20}{c}}
\bm s(k)  \\
\bm 0
\end{array} \right] \in \mathbb{C}^{2M}
\end{align}
and the $M$-dimensional filtered version of the expanded inputs from the convolution \eqref{eq:conv2} is calculated as
\begin{align}
\hat{\bm g}_i(k) &= [\hat g_i(kM+1),\hat g_i(kM+2),\dots,\hat g_i(kM+M)]^\mathrm{T} \nonumber\\
&=\mathrm{IFFT}[\bm g_i(\tilde k) \circ \bm s(\tilde k) ]~\text{last}~M~\text{elements}   .
\end{align}
At block $k$, let
\begin{align}
\hat{\bm y}(k) = \bm y(k)*s(k) = [\hat y(kM+1),\hat y(kM+2),\dots,\hat y(kM+M)]^\mathrm{T}
\end{align}
and the according residual noise vector is given by
\begin{align}
\bm e(k) = \bm d(k) - \hat{\bm y}(k) = \bm d(k) - \bm y(k)*s(k)   .
\end{align}

To solve the MSE minimization problem, the gradient estimates with respect to $\bm w_i(k)$ and $q(k)$ are derived as
\begin{align}\label{eq_wsdot_freq}
\frac{\partial J(k)}{\partial \bm w_i(k)}  = -2 \sum_{j=1}^M e(kM+j) \hat{\bm g}_i (kM+j)  = -2 \hat{\bm G}_i(k) \bm e(k)
\end{align}
and
\begin{align}
\frac{\partial J(k)}{\partial q(k)}  &= -2 \sum_{j=1}^M e(kM+j) \sum_{i=1}^{2P+1} \hat z_i (kM+j)  = -2\hat{\bm z}^\mathrm{T}(k) \bm e(k)
\end{align}
where $\hat{\bm z}(k) =\hat{\bm z}_1(k) + \hat{\bm z}_2(k) +\cdots + \hat{\bm z}_{2P+1}(k)$ with $\hat{\bm z}_i(k) = [\hat z_i(kM+1),\hat z_i(kM+2),\dots,\hat z_i(kM+M)]^\mathrm{T}$ and
\begin{align}\label{eq:zs_conv}
\hat z_i(kM+j) &= \hat{\bm h}_i^\mathrm{T}(kM+j)\bm w_i(k) \nonumber \\
&=\sum_{\ell=1}^M \hat h_i(kM+j-\ell+1) w_{iM-M+\ell}(k)   .
\end{align}
Recalling the overlap-save method to deal with the correlation of \eqref{eq_wsdot_freq} and the convolution of \eqref{eq:zs_conv}, the frequency domain implementations are expressed as
\begin{align}
\hat{\bm \phi}_i(k) =\mathrm{IFFT}[\bm e(\tilde k) \circ \mathrm{conj}(\hat{\bm g}_i(\tilde k) )]~\text{first}~M~\text{elements}
\end{align}
and
\begin{align}
\hat{\bm z}_i(k) = \mathrm{IFFT}[\hat{\bm h}_i(\tilde k) \circ \bm w_i(\tilde k) ]~\text{last}~M~\text{elements}
\end{align}
where we denote
\begin{align}
\hat{\bm g}_i(\tilde k) = \mathrm{FFT} \left[ \begin{array}{*{20}{c}}
\hat{\bm g}_i(k-1)  \\
\hat{\bm g}_i(k)
\end{array} \right]~\text{and}~
\hat{\bm h}_i(\tilde k) = \mathrm{FFT} \left[ \begin{array}{*{20}{c}}
\hat{\bm h}_i(k-1)  \\
\hat{\bm h}_i(k)
\end{array} \right]
\end{align}
with
\begin{align}
\hat{\bm h}_i(k) &= [\hat h_i(kM+1),\hat h_i(kM+2),\dots,\hat h_i(kM+M)]^\mathrm{T} \nonumber\\
&=\mathrm{IFFT}[\bm h_i(\tilde k) \circ \bm s(\tilde k) ]~\text{last}~M~\text{elements}   .
\end{align}

Then, by taking the SGD method, the learning rules are obtained as
\begin{align}
\bm w_i(\tilde k +1) &= \bm w_i(\tilde k) + \mu_w \mathrm{FFT} \left[ \begin{array}{*{20}{c}}
\hat{\bm \phi}_i(k)  \\
\bm 0
\end{array} \right] \nonumber\\
q(k+1) &= q(k) + \mu_q \hat{\bm z}^\mathrm{T}(k) \bm e(k)   .
\end{align}

As a consequence, the proposed FDEFsLMS algorithm is summarized in Table \ref{tab:FDEFsLMS}.
\begin{table}[!ht]\scriptsize
\centering
\renewcommand{\arraystretch}{1.2}
\setlength{\abovecaptionskip}{2pt}
\caption{Summary of the FDEFsLMS algorithm.}
\label{tab:FDEFsLMS}
\begin{tabular}{l}
\hline
\textbf{Initialization:} $\bm w(0), q(0), \mu_w, \mu_q$\\
\hline
  1:\hspace{6pt}\textbf{for} $k = 1,2,\dots$ \textbf{do}\\
  2:\hspace{12pt}\textbf{for} $i = 1,2,\dots,2P+1$ \textbf{do}\\
  3:\hspace{18pt}$\bm g_i(k) = [g_i(kM+1),g_i(kM+2),\dots,g_i(kM+M)]^\mathrm{T}$\\
  4:\hspace{18pt}$\bm g_i(\tilde k) = \mathrm{FFT} \left[ \begin{array}{*{20}{c}}
\bm g_i(k-1)  \\
\bm g_i(k)
\end{array} \right]$\\
  5:\hspace{18pt}$\bm y_i(k) =\mathrm{IFFT}[\bm g_i(\tilde k) \circ \bm w_i(\tilde k) ]$ last $M$ elements \\
  6:\hspace{12pt}\textbf{end for} \\
  7:\hspace{12pt}$\bm y(k) =\bm y_1(k) + \bm y_2(k) + \cdots + \bm y_{2P+1}(k)$\\
  8:\hspace{12pt}$\bm e(k) = \bm d(k) -\bm y(k)*s(k)$\\
  9:\hspace{12pt}$\bm e(\tilde k) = \mathrm{FFT} \left[ \begin{array}{*{20}{c}}
\bm 0  \\
\bm e(k)
\end{array} \right]$\\
  10:\hspace{8pt}\textbf{for} $i = 1,2,\dots,2P+1$ \textbf{do}\\
  11:\hspace{14pt}$\hat{\bm g}_i(k) =\mathrm{IFFT}[\bm g_i(\tilde k) \circ \bm s(\tilde k) ]$ last $M$ elements \\
  12:\hspace{14pt}$\hat{\bm g}_i(\tilde k) = \mathrm{FFT} \left[ \begin{array}{*{20}{c}}
\hat{\bm g}_i(k-1)  \\
\hat{\bm g}_i(k)
\end{array} \right]$\\
  13:\hspace{14pt}$ \hat{\bm \phi}_i(k) =\mathrm{IFFT}[\bm e(\tilde k) \circ \mathrm{conj}(\hat{\bm g}_i(\tilde k) )]$ first $M$ elements \\
  14:\hspace{14pt}$\bm w_i(\tilde k +1) = \bm w_i(\tilde k) + \mu_w \mathrm{FFT} \left[ \begin{array}{*{20}{c}}
\hat{\bm \phi}_i(k)  \\
\bm 0
\end{array} \right]$ \\
  15:\hspace{8pt}\textbf{end for} \\
  16:\hspace{8pt}\textbf{for} $i = 1,2,\dots,2P+1$ \textbf{do}\\
  17:\hspace{14pt}$\bm h_i(k) = [h_i(kM+1),h_i(kM+2),\dots,h_i(kM+M)]^\mathrm{T}$\\
  18:\hspace{14pt}$\bm h_i(\tilde k) = \mathrm{FFT} \left[ \begin{array}{*{20}{c}}
\bm h_i(k-1)  \\
\bm h_i(k)
\end{array} \right]$\\
  19:\hspace{14pt}$\hat{\bm h}_i(k) =\mathrm{IFFT}[\bm h_i(\tilde k) \circ \bm s(\tilde k) ]$ last $M$ elements \\
  20:\hspace{14pt}$\hat{\bm h}_i(\tilde k) = \mathrm{FFT} \left[ \begin{array}{*{20}{c}}
\hat{\bm h}_i(k-1)  \\
\hat{\bm h}_i(k)
\end{array} \right]$\\
  21:\hspace{14pt}$\hat{\bm z}_i(k) =\mathrm{IFFT}[\hat{\bm h}_i(\tilde k) \circ \bm w_i(\tilde k) ]$ last $M$ elements \\
  22:\hspace{8pt}\textbf{end for} \\
  23:\hspace{8pt}$\hat{\bm z}(k) =\hat{\bm z}_1(k) + \hat{\bm z}_2(k) + \cdots + \hat{\bm z}_{2P+1}(k)$\\
  24:\hspace{8pt}$q(k+1) = q(k) + \mu_q \hat{\bm z}^\mathrm{T}(k) \bm e(k)$\\
  25:\hspace{2pt}\textbf{end for} \\
\hline
\end{tabular}
\end{table}

\section{Performance analysis}

We introduce the stability, steady-state performance and computational complexity of the proposed frequency domain implementations in this section.

\subsection{Stability}

The stability as well as convergence of FDEFLN or FDEFsLMS is determined by the step sizes $\mu_w$ and $\mu_q$. Alternatively, we choose the FDEFLN algorithm to analyze the ranges of step sizes. The bounds on the step sizes will be derived in two aspects with respect to $\bm w_i(k)$ and $q(k)$, respectively. To ensure the convergence, the error signal $\bm e(k)$ for each block will gradually decrease during iterations. Hence, according to the energy conservation principle, we will have $\|\bm e(k+1)\|^2 \le \|\bm e(k)\|^2$ as the stability condition.

In the case of the first-order Taylor series expansion of $\|\bm e(k+1)\|^2$ at block $k$, we have
\begin{align}
\|\bm e(k+1)\|^2 = \|\bm e(k)\|^2 + \frac{\partial \|\bm e(k)\|^2}{\partial \bm w_i^\mathrm{T}(k)}\Delta\bm w_i(k)
\end{align}
where the higher-order terms are neglected, and $\frac{\partial \|\bm e(k)\|^2}{\partial \bm w_i^\mathrm{T}(k)} = \Big[\frac{\partial J(k)}{\partial \bm w_i(k)}\Big]^\mathrm{T}$ and $\Delta\bm w_i(k) = \bm w_i(k+1) - \bm w_i(k)$. We can further get
\begin{align}
\|\bm e(k+1)\|^2 &= \|\bm e(k)\|^2 + [-2\bm G_i(k) \bm e(k)]^\mathrm{T}[\mu_w \bm G_i(k) \bm e(k)] \nonumber\\
& = \|\bm e(k)\|^2 \cdot \Bigg[1 - 2\mu_w \frac{\|\bm G_i(k) \bm e(k)\|^2}{\|\bm e(k)\|^2}\Bigg]   .
\end{align}
In order to ensure the convergence, $\|\bm e(k+1)\|^2 \le \|\bm e(k)\|^2$ must be satisfied, and it yields
\begin{align}
0< 1 - 2\mu_w \frac{\|\bm G_i(k) \bm e(k)\|^2}{\|\bm e(k)\|^2}<1
\end{align}
which leads to
\begin{align}
0< \mu_w < \frac{\|\bm e(k)\|^2}{2\|\bm G_i(k) \bm e(k)\|^2}   .
\end{align}
For a frequency domain description, we thus consider
\begin{align}
\|\bm G_i(k) \bm e(k)\|^2 &\le \|\bm G_i(k)\|_\mathrm{F}^2 \cdot \|\bm e(k)\|^2 \nonumber\\
&< 2M[g_i^2(kM-M+1)+ \cdots + g_i^2(kM+M)] \cdot \|\bm e(k)\|^2  \nonumber\\
&= \|\bm g_i(\tilde k)\|^2 \cdot \|\bm e(k)\|^2
\end{align}
where $\|\cdot\|_\mathrm{F}$ stands for the Frobenius norm of its matrix argument, and we use the relation according to the Parseval's theorem \cite{Oppenheim2010Discrete}. Then, the bound on the step size $\mu_w$ as a stringent condition is given by
\begin{align}
0< \mu_w < \frac{1}{2\|\bm g_i(\tilde k)\|^2}  .
\end{align}

In a similar way, let us employ the first-order Taylor series expansion and $\Delta q(k) = q(k+1)-q(k)$, and we can obtain
\begin{align}
\|\bm e(k+1)\|^2 &= \|\bm e(k)\|^2 + \frac{\partial \|\bm e(k)\|^2}{\partial q(k)}\Delta q(k) \nonumber\\
&= \|\bm e(k)\|^2 + [-2\bm z^\mathrm{T}(k) \bm e(k)] [\mu_q \bm z^\mathrm{T}(k) \bm e(k)] \nonumber\\
& = \|\bm e(k)\|^2 \cdot \Bigg[1 - 2\mu_q \frac{\|\bm z^\mathrm{T}(k) \bm e(k)\|^2}{\|\bm e(k)\|^2}\Bigg]   .
\end{align}
Consider the following relation
\begin{align}
\|\bm z^\mathrm{T}(k) \bm e(k)\|^2 &\le \|\bm z(k)\|^2 \cdot \|\bm e(k)\|^2 \nonumber\\
&\le [\|\bm z_1(k)\|^2 + \cdots + \|\bm z_{2P+1}(k)\|^2] \cdot \|\bm e(k)\|^2  \nonumber\\
&\le (2P+1) \max_{\scriptscriptstyle 1\le i \le 2P+1}(\|\bm z_i(k)\|^2) \cdot \|\bm e(k)\|^2  .
\end{align}
The bound on the step size $\mu_q$ can be obtained by
\begin{align}
0< \mu_q < \frac{1}{2(2P+1) \max_{\scriptscriptstyle 1\le i \le 2P+1}(\|\bm z_i(k)\|^2)}   .
\end{align}

\subsection{Steady-state performance}

We present a statistical analysis of FDEFLN in terms of the mean-square performance. To proceed with the mathematical analysis, the following assumptions are made.

\emph{Assumption 1:} The i.i.d. noise signal $\eta(n)$ is a Gaussian sequence with zero-mean and variance $\sigma_\eta^2$, and is independent of $\bm u(n), \bm w (n)$ and $q(n)$.

\emph{Assumption 2:} $\bm u(n), \bm w (n)$ and $q(n)$ are mutually statistically independent. This is the well-known independence assumption.

\emph{Assumption 3:} The error sequence $e(n)$ is asymptotically uncorrelated with $\|\bm g(n)\|^2$ and $|\bm h^\mathrm{T} \bm w(n)|^2$. This has been commonly used for analyzing EFLN-based algorithms at steady-state \cite{Patel2021Convergence,Yu2021RobustAdaptive}.

According to the literature in \cite{Yang2019Mean}, we first present the time domain counterparts for the learning rules \eqref{eq:wrule} and \eqref{eq:qrule} of the weight vector and exponential factor, which allows us to make the analysis tractable. Let $\bm {\mathcal{F}} \in \mathbb{C}^{2M \times 2M}$ denote the DFT matrix. The frequency domain expanded input diagonal matrix is introduced as $\bm G_{fi}(\tilde k) = \mathrm{diag}\{\bm g_i(\tilde k)\} \in \mathbb{C}^{2M \times 2M}$. The learning rule \eqref{eq:wrule} with regard to the weight vector can be rewritten as the frequency domain matrix form
\begin{align}\label{eq:Ehw1}
\bm{w}_i( \tilde k + 1 ) = \bm {\mathcal{F}} \bm{Q}_{10} \bm {\mathcal{F}}^{-1}[ \bm{w}_i( \tilde k ) + {\mu _w}\bm{G}_{fi}^\mathrm{H}( \tilde k )\bm{E}( \tilde k ) ]
\end{align}
where the superscript $(\cdot)^\mathrm{H}$ denotes the Hermitian operation, and $\bm{Q}_{10} = \Big[\begin{smallmatrix}
\bm{I}_M & \bm{0} \\
\bm{0} & \bm{0}
\end{smallmatrix} \Big] \in \mathbb{R}^{2M \times 2M}$ is a constraint matrix that forces the last $M$ elements to zero. We introduce the circular matrix results as
\begin{align}\label{eq:Ehw3}
\bm{G}_{fi}( \tilde k ) &= \bm {\mathcal{F}} \bm{G}_{ci}( k )\bm {\mathcal{F}}^{ - 1} \nonumber\\
\bm{G}_{fi}^\mathrm{H}( \tilde k ) &= \bm {\mathcal{F}} \bm{G}_{ci}^\mathrm{T}( k )\bm {\mathcal{F}}^{ - 1}
\end{align}
where $\bm{G}_{ci}( k ) \in \mathbb{R}^{2M \times 2M} $ is the circular matrix, whose first column is $[ {g_i}( kM - M + 1 ), \dots ,{g_i}( kM + M ) ]^\mathrm{T}$, and the other columns are obtained by circularly shifting the previous column by one element \cite{Yang2019Mean}. The matrix $\bm{G}_{ci}( k )$ can also be written in a block matrix form as
\begin{align}
\bm{G}_{ci}( k ) = \left[ {\begin{array}{*{20}{c}}
{\bm{G}_{ai}( k )}&{\bm{G}_{bi}( k )}\\
{\bm{G}_{bi}( k )}&{\bm{G}_{ai}( k )}
\end{array}} \right]
\end{align}
where we can easily get $\bm{G}_{bi}( k ) = \bm G_i^\mathrm{T}(k) \in \mathbb{R}^{M \times M}$. Pre-multiplying both sides of \eqref{eq:Ehw1} by $\bm {\mathcal{F}}^{ - 1}$ and using the circular matrix results \eqref{eq:Ehw3}, we have
\begin{align}
\left[ \begin{array}{*{20}{c}}
{\bm{w}_i( k + 1 )}\\
{\bm{0}}
\end{array} \right] &= \left[ \begin{array}{*{20}{c}}
{\bm{w}_i( k )}\\
\bm{0}
\end{array} \right] + {\mu _w}\bm {\mathcal{F}}^{ - 1}\bm {\mathcal{F}}\bm{Q}_{10}\bm {\mathcal{F}}^{ - 1}\bm{G}_{fi}^\mathrm{H}( k )\bm {\mathcal{F}}\left[ \begin{array}{*{20}{c}}
\bm{0}\\
\bm{e}( k )
\end{array} \right] \nonumber\\
 &= \left[ \begin{array}{*{20}{c}}
\bm{w}_i( k )\\
\bm{0}
\end{array} \right] + {\mu _w}\bm{Q}_{10}{\left[ \begin{array}{*{20}{c}}
{\bm{G}_{ai}( k )}&{\bm{G}_{bi}( k )}\\
{\bm{G}_{bi}( k )}&{\bm{G}_{ai}( k )}
\end{array} \right]^\mathrm{H}}\left[ \begin{array}{*{20}{c}}
\bm{0}\\
\bm{e}( k )
\end{array} \right] \nonumber\\
 &= \left[ {\begin{array}{*{20}{c}}
\bm{w}_i( k )\\
\bm{0}
\end{array}} \right ] + \left[ \begin{array}{*{20}{c}}
{\mu _w}\bm{G}_{i}( k )\bm{e}( k )\\
\bm{0}
\end{array} \right ].
\end{align}
Hence, the updating equation of the weight vector $\bm w_i(k)$ can be formulated as
\begin{align}
\bm{w}_i( k + 1 ) = \bm{w}_i( k ) + {\mu _w}\bm{G}_{i}( k )\bm{e}( k ).
\end{align}
We can directly obtain the corresponding learning rule for the weight vector $\bm w(k)$ as
\begin{align}\label{eq:wfin}
\bm{w}( k + 1 ) = \bm{w}( k ) + {\mu _w}\bm{G}( k )\bm{e}( k ).
\end{align}

The filtered output vector has been given by $\bm y(k) = \sum_{i = 1}^{2P + 1} \bm{G}_{i}^\mathrm{T}( k )\bm{w}_i( k ) = \bm G^\mathrm{T}(k) \bm w(k)$. We denote the diagonal matrix $\bm H_{fi}(\tilde k) = \mathrm{diag}\{\bm h_i(\tilde k)\} \in \mathbb{C}^{2M \times 2M}$, and we can derive
\begin{align}
\left[ {\begin{array}{*{20}{c}}
{\bm{0}}\\
\bm{z}_i( k )
\end{array}} \right] &= \bm{Q}_{01}\bm {\mathcal{F}}^{ - 1}\bm{H}_{fi}( \tilde k )\bm{w}_{i}( \tilde k ) \nonumber \\
& = \left[ {\begin{array}{*{20}{c}}
{\bm{0}}\\
\bm{H}_i^\mathrm{T}( k ) \bm w_i(k)
\end{array}} \right]
\end{align}
where $\bm{Q}_{01} = \Big[\begin{smallmatrix}
\bm{0}& \bm{0} \\
\bm{0} & \bm{I}_M
\end{smallmatrix} \Big] \in \mathbb{R}^{2M \times 2M}$ is a constraint matrix that forces the first $M$ elements to zero. We can get $\bm{z}_i( k ) = \bm{H}_i^\mathrm{T}( k ) \bm w_i(k)$ and further obtain $\bm z(k) = \sum_{i = 1}^{2P + 1} \bm{H}_{i}^\mathrm{T}( k )\bm{w}_i( k ) = \bm H^\mathrm{T}(k) \bm w(k)$. Therefore, the updating equation of the exponential factor $q(k)$ can be formulated as
\begin{align}\label{eq:qfin}
q( k + 1 ) = q( k ) + {\mu _q}\bm{w}^\mathrm{T}( k ) \bm{H}( k )\bm{e}( k ).
\end{align}

The steady-state performance analysis of FDEFLN will be discussed in the mean-square sense. We introduce the \emph{a priori error} vector as
\begin{align}\label{eq:Ehw9}
\bm \varepsilon ( k ) &= \bar{\bm y }( k ) - \bm y ( k ) \nonumber\\
 &= \bar {\bm G}^\mathrm{T}(k) \bar{\bm w} - \bm G ( k )\bm w ( k )  \nonumber\\
 &= \bar {\bm G}^\mathrm{T}(k)[\bar{\bm w} - \bm w ( k ) ] + [ \bar {\bm G}( k ) - \bm G ( k ) ]^\mathrm{T} \bm w ( k )  \nonumber\\
 &= {\bm \varepsilon}_w( k ) + {\bm \varepsilon }_q( k )
\end{align}
where $\bar{\bm y }( k ) = [\bar y(kM+1),\bar y(kM+2),\dots,\bar y(kM+M)]^\mathrm{T}$ and $\bar {\bm G}(n)$ is the expanded input matrix with the ideal exponential factor $\bar q$, and we denote ${\bm \varepsilon}_w( k ) = \bar {\bm G}^\mathrm{T}(k)\tilde{\bm w}( k )$ and ${\bm \varepsilon }_q( k ) = \tilde {\bm G}^\mathrm{T}( k )\bm w ( k )$ with $\tilde{\bm w}( k ) = \bar{\bm w} - \bm w ( k )$ and $\tilde {\bm G}( k ) = \bar {\bm G}( k ) - \bm G ( k )$ being the weight vector error and the expanded input matrix error, respectively.

As a performance metric, the steady-state excess mean-square error (EMSE) will be evaluated. Concerning the steady-state performance for $k \to \infty$, we deduce the theoretical steady-state EMSE
\begin{align}\label{eq:EMSESS}
\mathrm{EMSE}(\infty) &= \frac{1}{M} \lim_{k \to \infty} \mathbb{E}\{\|\bm \varepsilon ( k ) \|^2\}  \nonumber\\
&=\frac{1}{M} \lim_{k \to \infty} \mathbb{E}\{\| {\bm \varepsilon}_w( k ) + {\bm \varepsilon }_q( k ) \|^2\}
\end{align}
where $\mathbb{E}\{\cdot\}$ is the expectation operation.

Subtracting both sides of \eqref{eq:wfin} from $\bar {\bm w}$ yields
\begin{align}\label{eq:ssw}
\tilde{\bm w} (k + 1) = \tilde{\bm w} (k) - \mu_w \bm G(k) \bm e(k)    .
\end{align}
Calculating the $l_2$-norm square of both sides of \eqref{eq:ssw} and taking the expectation operation, we have
\begin{align}
\mathbb{E}\{ \|\tilde{\bm w} (k + 1)\|^2 \} = \mathbb{E}\{ \|\tilde{\bm w} (k)\|^2 \}- 2 \mu_w \mathbb{E}\{\bm e^\mathrm{T}(k)  \bm G^\mathrm{T}(k) \tilde{\bm w} (k)\} + \mu_w^2 \mathbb{E}\{\|\bm G(k) \bm e(k)\|^2\}  .
\end{align}
It is noted that $\lim_{k \to \infty}\mathbb{E}\{ \|\tilde{\bm w} (k + 1)\|^2 \} =\lim_{k \to \infty}\mathbb{E}\{ \|\tilde{\bm w} (k)\|^2 \}$ holds, thereby leading to
\begin{align}\label{eq:leandriw}
2 \mathbb{E}\{\bm e^\mathrm{T}(k)  \bm G^\mathrm{T}(k) \tilde{\bm w} (k)\} = \mu_w \mathbb{E}\{\|\bm G(k) \bm e(k)\|^2\} .
\end{align}

Assuming $q(k) \to \bar q$ for $k \to \infty$, we have $\bm e(k) \approx {\bm \varepsilon}_w(k) + \bm \eta(k)$, where $\bm \eta( k ) = [\eta(kM+1),\eta(kM+2),\dots,\eta(kM+M)]^\mathrm{T}$. According to Assumptions 1-3, we compute the left-hand side of \eqref{eq:leandriw} as
\begin{align}\label{eq:lew}
2 \mathbb{E} \{  \bm e^\mathrm{T}(k)  \bm G^\mathrm{T}(k) \tilde{\bm w} (k)\} &= 2 \mathbb{E}\{ [{\bm \varepsilon}_w(k) + \bm \eta(k)]^\mathrm{T}  \bm G^\mathrm{T}(k) \tilde{\bm w} (k) \}   \nonumber\\
&= 2 \mathbb{E}\{  {\bm \varepsilon}_w^\mathrm{T}(k)  \bm G^\mathrm{T}(k) \tilde{\bm w} (k)\} = 2 \mathbb{E}\{ \| {\bm \varepsilon}_w(k)\|^2\}
\end{align}
and we compute the right-hand side of \eqref{eq:leandriw} as
\begin{align}
\mu_w \mathbb{E}\{\|\bm G(k) \bm e(k)\|^2\} &= \mu_w \mathbb{E}\{\|\bm G(k) {\bm \varepsilon}_w(k) + \bm G(k) \bm \eta(k)\|^2\} \nonumber\\
&=\mu_w \mathbb{E}\{\|\bm G(k) {\bm \varepsilon}_w(k)\|^2\} + \mu_w \mathbb{E}\{\|\bm G(k) \bm \eta(k)\|^2\} \nonumber\\
&=\mu_w \mathrm{Tr} [ \mathbb{E}\{ {\bm \varepsilon}_w(k) {\bm \varepsilon}_w^\mathrm{T}(k)\} \mathbb{E}\{ \bm G^\mathrm{T}(k)\bm G(k)\}  ] + \mu_w \mathrm{Tr} [ \mathbb{E}\{ \bm \eta(k)  {\bm \eta}^\mathrm{T}(k)\} \mathbb{E}\{ \bm G^\mathrm{T}(k)\bm G(k)\}  ]
\end{align}
where $\mathrm{Tr}[\cdot]$ represents the matrix trace. Consider the results of $\mathbb{E}\{ {\bm \varepsilon}_w(k) {\bm \varepsilon}_w^\mathrm{T}(k)\} = \frac{1}{M}\mathbb{E}\{ \| {\bm \varepsilon}_w(k)\|^2\} \bm S$ with all elements of $\bm S \in \mathbb{R}^{M \times M}$ being ones, and $\mathbb{E}\{ \bm \eta(k)  {\bm \eta}^\mathrm{T}(k)\} = \sigma_\eta^2 \bm I_M$. We can further get
\begin{align}\label{eq:riw}
\mu_w \mathbb{E}\{\|\bm G(k) \bm e(k)\|^2\} = \mu_w \frac{1}{M}\mathbb{E}\{ \| {\bm \varepsilon}_w(k)\|^2\} \mathrm{Tr}  [ \bm S\mathbb{E}\{ \bm G^\mathrm{T}(k) \bm G(k)\}  ] + \mu_w \sigma_\eta^2\mathrm{Tr} [ \mathbb{E}\{ \bm G^\mathrm{T}(k) \bm G(k)\}  ]  .
\end{align}
Plugging \eqref{eq:lew} and \eqref{eq:riw} into \eqref{eq:leandriw}, and considering the steady-state for $k \to \infty$, we have
\begin{align}\label{eq:winf}
\lim_{k \to \infty} \mathbb{E}\{\| {\bm \varepsilon}_w(k)\|^2\} = \frac{\mu_w M \sigma_\eta^2 \mathrm{Tr} [ \mathbb{E}\{ \bm G^\mathrm{T}(k)\bm G(k) \}  ]}{2M-\mu_w \mathrm{Tr}  [ \bm S\mathbb{E}\{ \bm G^\mathrm{T}(k)\bm G(k)\}  ]}   .
\end{align}

Subtracting both sides of \eqref{eq:qfin} from $\bar q$, we have
\begin{align}\label{eq:ssq}
\tilde q (k + 1) = \tilde q (k) - \mu_q  \bm w^\mathrm{T}(k) \bm H(k) \bm e(k) .
\end{align}
Calculating the $l_2$-norm square of both sides of \eqref{eq:ssq} and taking the mathematical expectation, we obtain
\begin{align}
\mathbb{E}\{ \tilde q^2 (k + 1) \} = \mathbb{E}\{ \tilde q^2 (k) \} - 2 \mu_q \mathbb{E}\{ \tilde q (k) \bm w^\mathrm{T}(k) \bm H(k) \bm e(k) \} + \mu_q^2 \mathbb{E}\{ |\bm w^\mathrm{T}(k) \bm H(k) \bm e(k)|^2 \}  .
\end{align}
Also noting that $\lim_{k \to \infty}\mathbb{E}\{ \tilde q^2 (k + 1) \} =\lim_{k \to \infty}\mathbb{E}\{ \tilde q^2 (k + 1) \}$ holds, it yields
\begin{align}\label{eq:leandriq}
2 \mathbb{E}\{ \tilde q (k) \bm w^\mathrm{T}(k) \bm H(k) \bm e(k) \} = \mu_q \mathbb{E}\{ |\bm w^\mathrm{T}(k) \bm H(k) \bm e(k)|^2 \}  .
\end{align}
In the following, it will be proved that $\bar {\bm G}( k ) - \bm G ( k ) \approx [\bar q - q(k)]\bm H ( k )$ holds. The entries of $\bar {\bm G}( k ) - \bm G ( k )$ for each $i = 1,2,\dots,2P+1$ can be obtained as
\begin{align}
\bar{\bm g}_i(k) - \bm g_i(k) &=
\left\{\begin{aligned}
&\bm 0,  & &  i = 1\\
&[e^{-\bar q|\bm u(k)|} - e^{-q(k)|\bm u(k)|}] \circ \sin[p \pi \bm u(k)], & &  i = 2p\\
&[e^{-\bar q|\bm u(k)|} - e^{-q(k)|\bm u(k)|}] \circ \cos[p \pi \bm u(k)], & &  i = 2p+1
\end{aligned}\right. \nonumber\\
&=\left\{\begin{aligned}
&\bm 0,  & &  i = 1\\
&[e^{-[\bar q - q(k)] |\bm u(k)|} - 1] \circ e^{-q(k)|\bm u(k)|} \circ \sin[p \pi \bm u(k)], & &  i = 2p\\
&[e^{-[\bar q - q(k)] |\bm u(k)|} - 1] \circ e^{-q(k)|\bm u(k)|} \circ \cos[p \pi \bm u(k)], & &  i = 2p+1   .
\end{aligned}\right.
\end{align}
Considering the Taylor series approximation and neglecting the higher-order terms, it yields $e^{-[\bar q - q(k)] |\bm u(k)|} - 1 \approx -[\bar q - q(k)] |\bm u(k)|$. We can further get $\bar{\bm g}_i(k) - \bm g_i(k) \approx -[\bar q - q(k)] \bm h_i(k)$, and $\bar {\bm G}( k ) - \bm G ( k ) \approx [\bar q - q(k)]\bm H ( k )$ is confirmed, i.e., $\tilde {\bm G}( k ) \approx \tilde q(k)\bm H ( k )$. Obviously, ${\bm \varepsilon}_q(k) \approx \tilde q(k)\bm H^\mathrm{T} ( k ) \bm w(k)$ holds.

Assuming $\bm w(k) \to \bar {\bm w}$ for $k \to \infty$, then $\bm e(k) \approx {\bm \varepsilon}_q(k) + \bm \eta(k)$, as well as Assumptions 1-3, both sides of \eqref{eq:leandriq} can be calculated as
\begin{align}\label{eq:leq}
2 \mathbb{E}\{ \tilde q (k) \bm w^\mathrm{T}(k) \bm H(k) \bm e(k) \} &= 2 \mathbb{E}\{ \tilde q (k) \bm w^\mathrm{T}(k) \bm H(k) \bm [{\bm \varepsilon}_q(k) + \bm \eta(k)] \} \nonumber\\
&= 2 \mathbb{E}\{ \tilde q (k) \bm w^\mathrm{T}(k) \bm H(k) {\bm \varepsilon}_q(k)\} = 2 \mathbb{E}\{ \| {\bm \varepsilon}_q(k)\|^2\}
\end{align}
and
\begin{align}\label{eq:riq}
&~\mu_q \mathbb{E}\{ |\bm w^\mathrm{T}(k) \bm H(k) \bm e(k)|^2 \} \nonumber\\
=&~ \mu_q \mathbb{E}\{ |\bm w^\mathrm{T}(k) \bm H(k) {\bm \varepsilon}_q(k)|^2 \} + \mu_q \mathbb{E}\{ |\bm w^\mathrm{T}(k) \bm H(k) \bm \eta(k)|^2 \}  \nonumber\\
=&~ \mu_q \frac{1}{M}\mathbb{E}\{ \| {\bm \varepsilon}_q(k)\|^2\} \mathrm{Tr}  [ \bm S\mathbb{E}\{ \bm H^\mathrm{T}(k)\bm w(k) \bm w^\mathrm{T}(k)\bm H(k)\}  ] + \mu_q \sigma_\eta^2\mathrm{Tr} [ \mathbb{E}\{ \bm H^\mathrm{T}(k)\bm w(k) \bm w^\mathrm{T}(k)\bm H(k)\}   ]  .
\end{align}
Substituting \eqref{eq:leq} and \eqref{eq:riq} into \eqref{eq:leandriq} and evaluating them at steady-state, we have
\begin{align}\label{eq:qinf}
\lim_{k \to \infty} \mathbb{E}\{\| {\bm \varepsilon}_q(k)\|^2\} = \frac{\mu_q M \sigma_\eta^2 \mathrm{Tr} [ \mathbb{E}\{ \bm H^\mathrm{T}(k)\bm w(k) \bm w^\mathrm{T}(k)\bm H(k)\}  ]}{2M-\mu_q \mathrm{Tr}  [ \bm S\mathbb{E}\{ \bm H^\mathrm{T}(k)\bm w(k) \bm w^\mathrm{T}(k)\bm H(k)\}   ]}   .
\end{align}

Since $\bm w(k) \to \bar{\bm w}$ and $q(k) \to \bar q$ for $k \to \infty$, it implies that
\begin{align}
\lim_{k \to \infty} \mathbb{E}\{ {\bm \varepsilon}_w^\mathrm{T}(k) {\bm \varepsilon}_q(k) \}  =  \mathrm{Tr} [\mathbb{E}\{ \bm w(k) \tilde{\bm w}^\mathrm{T}(k) \}  \mathbb{E}\{ \tilde q(k) \bm H(k) \bar {\bm G}^\mathrm{T}(k) \}  ] = 0  .
\end{align}
Inserting \eqref{eq:winf} and \eqref{eq:qinf} into \eqref{eq:EMSESS}, the steady-state EMSE is then evaluated as
\begin{align}\label{eq:ssEMSE}
\mathrm{EMSE}(\infty) &= \frac{1}{M} \lim_{k \to \infty} \mathbb{E}\{\|\bm \varepsilon ( k ) \|^2\}  \nonumber\\
&=\frac{1}{M} \lim_{k \to \infty} \mathbb{E}\{\| {\bm \varepsilon}_w( k )\|^2\} + \frac{1}{M} \lim_{k \to \infty} \mathbb{E}\{\| {\bm \varepsilon}_q( k )\|^2\} \nonumber\\
& = \frac{\mu_w  \sigma_\eta^2 \mathrm{Tr} [ \mathbb{E}\{ \bm G^\mathrm{T}(k)\bm G(k) \}  ]}{2M-\mu_w \mathrm{Tr}  [ \bm S\mathbb{E}\{ \bm G^\mathrm{T}(k)\bm G(k)\}  ]} + \frac{\mu_q  \sigma_\eta^2 \mathrm{Tr} [ \mathbb{E}\{ \bm H^\mathrm{T}(k)\bm w(k) \bm w^\mathrm{T}(k)\bm H(k)\}  ]}{2M-\mu_q \mathrm{Tr}  [ \bm S\mathbb{E}\{ \bm H^\mathrm{T}(k)\bm w(k) \bm w^\mathrm{T}(k)\bm H(k)\}   ]}  .
\end{align}
As a consequence, a closed-form expression of the theoretical steady-state value of FDEFLN is given by \eqref{eq:ssEMSE}.

\subsection{Computational complexity}

We evaluate the computational requirements of the proposed FDEFLN and FDEFsLMS algorithms as compared to the EFLN and EFsLMS algorithms. The total numbers of multiplications and additions are computed for each iteration in a data block, and all computations are carried out in real-valued arithmetic. In detail, Table \ref{tab:complexity} summarizes the total computational complexities of adaptive algorithms.

\begin{figure}[!ht]
\centering
\includegraphics[width=2.7in]{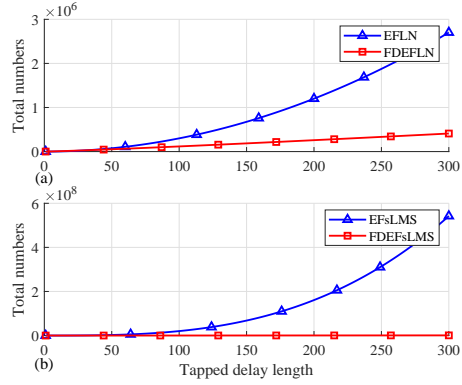}
\caption{Total numbers of multiplications and additions versus $M$, with $P = 2$ and $N = M$. (a) For the EFLN and FDEFLN algorithms. (b) For the EFsLMS and FDEFsLMS algorithms.}
\label{fig:computational_complexity}
\end{figure}

The proposed FDEFLN and FDEFsLMS algorithms execute the filtering and adaptation procedures in frequency domain, so they can reduce the computational requirement for each iteration in a data block. Fig. \ref{fig:computational_complexity} shows the total numbers of multiplications and additions with respect to tapped delay length $M$ for adaptive algorithms, considering $P = 2$ and $N = M$. It is evident that the FDEFLN and FDEFsLMS algorithms have the remarkable reduction of computational requirement in comparison with the corresponding EFLN and EFsLMS algorithms. Moreover, we also observe that the frequency domain algorithms can heavily decrease the computational cost with the increase of $M$.
\begin{table*}[!ht]
\centering
\renewcommand{\arraystretch}{1.3}
\setlength{\abovecaptionskip}{2pt}
\caption{Summary of the computational complexities for adaptive algorithms.}
\label{tab:complexity}
\resizebox{\linewidth}{!}{
\begin{tabular}{|c|c|c|c|c|}
\hline
\multirow{2}{*}{\makecell{Algorithms}} & \multicolumn{2}{c|}{EFLN}                       & \multicolumn{2}{c|}{FDEFLN}             \\ \cline{2-5}
                                       & Multiplications           & Additions           & Multiplications          & Additions    \\ \hline
Filtering         & $M(2P+1)$        & $M(2P+1)-1$   & $(4M\log_2 2M + 8M)(2P+1)$                         & $(4M\log_2 2M + 4M)(2P+1)+2MP$          \\ \hline
Error             & -                & 1             & $2M\log_2 2M$                                      & $2M\log_2 2M+M$                         \\ \hline
Weight adaptation & $M(2P+1)+1$      & $M(2P+1)$     & $(4M\log_2 2M + 10M)(2P+1)$                        & $(4M\log_2 2M + 8M)(2P+1)$              \\ \hline
Factor adaptation & $M(2P+1)+2$      & $M(2P+1)$     & $(4M\log_2 2M + 8M)(2P+1)+M+1$                     & $(4M\log_2 2M + 5M)(2P+1)$              \\ \hline
Total operations  & $3M^2(2P+1)+3M$  & $3M^2(2P+1)$  & $(12M\log_2 2M + 26M)(2P+1) + 2M\log_2 2M + M +1$  & $(12M\log_2 2M + 18M)(2P+1) + 2M\log_2 2M$ \\ \hline
\hline
\multirow{2}{*}{\makecell{Algorithms}} & \multicolumn{2}{c|}{EFsLMS}                     & \multicolumn{2}{c|}{FDEFsLMS}           \\ \cline{2-5}
                                       & Multiplications           & Additions           & Multiplications          & Additions    \\ \hline
Filtering         & $M(2P+1)$        & $M(2P+1)-1$   & $(4M\log_2 2M + 8M)(2P+1)$                         & $(4M\log_2 2M + 4M)(2P+1)+2MP$          \\ \hline
Error             & $N$              & $N$           & $2M\log_2 2M+MN$                                   & $2M\log_2 2M+MN$                        \\ \hline
Weight adaptation & $M(2P+1)(N+1)+1$ & $MN(2P+1)$    & $(8M\log_2 2M + 18M)(2P+1)$                        & $(8M\log_2 2M + 12M)(2P+1)$             \\ \hline
Factor adaptation & $M(2P+1)(N+1)+2$ & $MN(2P+1)$    & $(8M\log_2 2M + 16M)(2P+1)+M+1$                    & $(8M\log_2 2M + 9M)(2P+1)$              \\ \hline
Total operations  & $M^2(2P+1)(2N+3)+MN+3M$ & $M^2(2P+1)(2N+1)+MN-M$ & $(20M\log_2 2M + 42M)(2P+1) + 2M\log_2 2M + MN+M +1$ & $(20M\log_2 2M + 25M)(2P+1) + 2M\log_2 2M +2MP+MN$    \\ \hline
\end{tabular}  }
\end{table*}

\section{Numerical results}

In this section, illustrative examples with applications of NSI, NAEC and NANC are performed to verify the proposed algorithms. We mainly evaluate the FDEFLN and FDEFsLMS algorithms in frequency domain implementation, and compare their computational time with the corresponding time domain algorithms. All algorithms are edited in MATLAB R2018a and implemented in a computer with the 2.9GHz Intel Core i5-9400F CPU and 24GB RAM.

\subsection{Verification of performance analysis}

Consider that an unknown nonlinear system comprises an EFLN nonlinear expansion block and a long FIR filter. The identified impulse response is taken as the FIR filter weights $\bm w_\mathrm{o}$ with length 1000, which is generated by a seventh-order system with the transfer function as
\begin{align}
G(\mathrm{s}) =  \frac{\mathrm{-2.8e12s^3+4.6e18s^2+6.4e21s+3.2e27}}{\mathrm{s^7+1e4s^6+2.6e9s^5+1.2e13s^4+1.2e18s^3+2.1e21s^2+9.4e23s+9.7e26}}  .
\end{align}
The ideal exponential factor is assumed to be $q_\mathrm{o} = -0.4$. This identified model with relative long filter weights has been typically used for system identification purposes \cite{Yang2020Frequency}.

The input $u(n)$ is extracted from a uniform distribution over $[-1,1]$. The desired signal is contaminated by a zero-mean Gaussian noise with a signal-to-noise ratio (SNR) of 40dB. The block length and the tapped delay length are equally set as $M=200$, and the functional expansion order is taken as $P=2$. All frequency domain weights and the exponential factor are initialized as zero elements. The step sizes are chosen as $\mu_w = 0.0002$ and $\mu_q = 0.0005$ in this identification process.

\begin{figure}[!t]
\centering
\includegraphics[width=2.7in]{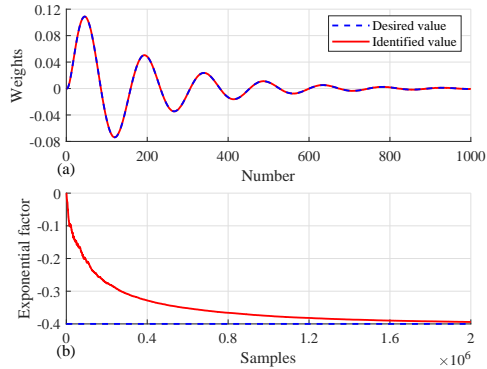}
\caption{Convergence result of the FDEFLN algorithm. (a) Long filter weights. (b) Variations of the exponential factor.}
\label{fig:iden_result}
\end{figure}
\begin{figure}[!t]
\centering
\includegraphics[width=2.7in]{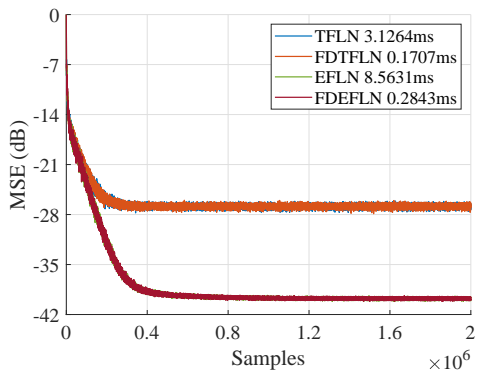}
\caption{MSE curves of adaptive algorithms, with $\mu_w = 0.0002,\mu_q = 0.0005$ for FDEFLN and EFLN, and $\mu = 0.0002$ for TFLN and FDTFLN.}
\label{fig:comparison_iden}
\end{figure}
The identification result of the FDEFLN algorithm is shown in Fig. \ref{fig:iden_result}. This indicates that the proposed FDEFLN algorithm can exactly converge to the desired weights and the exponential factor, so it achieves a precise system identification. We further compare the performance of FDEFLN with that of EFLN and the traditional TFLN and frequency domain TFLN (FDTFLN) algorithms in the same simulation parameters, where the $\mathrm{MSE}=10\log_{10}\mathbb{E}\{e^2(n)\}$ is utilized as the performance metric. The MSE curves of adaptive algorithms are illustrated in Fig. \ref{fig:comparison_iden}. We can see that the MSE evolutions of the FDEFLN and EFLN algorithms overlap very well, which implies that they have the equivalent convergence property, and have lower MSE level as compared to the traditional TFLN-based algorithms. Additionally, the average execution time for $M$ samples has been measured. It is worth noting that FDEFLN consumes 0.2843ms computational time for each data block, while EFLN consumes 8.5631ms for $M$ samples. Thanks to the frequency domain implementation, the proposed FDEFLN algorithm will possess much lower computational complexity.

\begin{figure}[!t]
\centering
\includegraphics[width=2.7in]{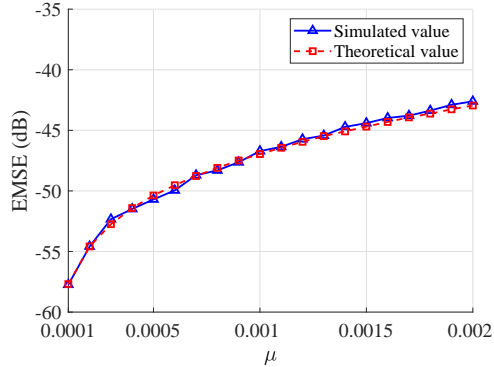}
\caption{Simulated and theoretical values of steady-state EMSE versus step sizes $\mu = 0.0001,0.0002,\dots,0.002$.}
\label{fig:theoretical_EMSE}
\end{figure}

Considering the nonlinear system described above, we further validate the theoretical steady-state result of FDEFLN. The input $u(n)$ is drawn from a Gaussian signal with zero-mean and unit variance. The step sizes are taken as $\mu_w = \mu_q = \mu$ with the range of $[0.0001,0.002]$, and the block length is set to $M = 64$. The expectations involved in \eqref{eq:ssEMSE} of the theoretical value are computed by using the average of the last $50M$ samples over 100 independent trials. The simulated value is also obtained by averaging over 100 independent trials. From Fig. \ref{fig:theoretical_EMSE}, we can see the verification of the simulated and theoretical steady-state EMSE versus step sizes, where a good agreement between the simulated and theoretical values is exhibited.

\subsection{Implementations of FDEFLN}

\subsubsection{In the NSI scenario}

\begin{figure}[!ht]
\centering
\includegraphics[width=2.6in]{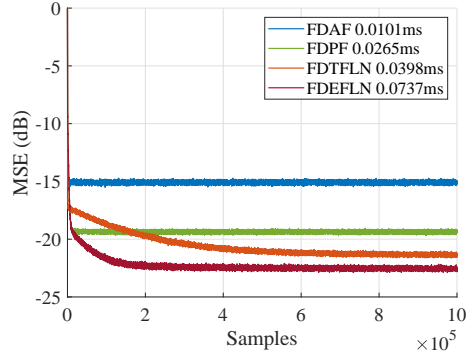}
\caption{Comparison of adaptive algorithms in the NSI scenario, with $M=32$, $\mu_w = 0.003,\mu_q = 0.05$ for FDEFLN, $\mu = 0.003$ for FDTFLN and FDPF, and $\mu = 0.09$ for FDAF.}
\label{fig:comparison_NSI}
\end{figure}

Consider a nonlinear system with the input-output characteristic given by
\begin{align}
\bar y(n) = 0.6\sin^3[\pi u(n)] - \frac{2}{u^3(n) + 2} - 0.1\cos[4\pi u(n-4)] +1.125
\end{align}
in a typical NSI scenario \cite{Patel2016Design}, and $\bar y(n)$ is contaminated by a Gaussian white noise with a SNR of 40dB. The input $u(n)$ is employed by a uniformly distributed signal over $[-0.5,0.5]$. In this NSI task, the performance of FDEFLN is compared with that obtained by the linear frequency domain adaptive filtering (FDAF) and the nonlinear FDPF and FDTFLN algorithms. The simulation parameters of all algorithms are set to obtain the same initial convergence. Fig. \ref{fig:comparison_NSI} shows the comparison of these frequency domain algorithms. We can see the performance benefit of the nonlinear algorithms compared to the linear FDAF algorithm. In addition, the proposed FDEFLN algorithm has much smaller steady-state misalignment than other nonlinear frequency domain approaches.

\subsubsection{In the NAEC scenario}

\begin{figure}[!t]
\centering
\includegraphics[width=2.6in]{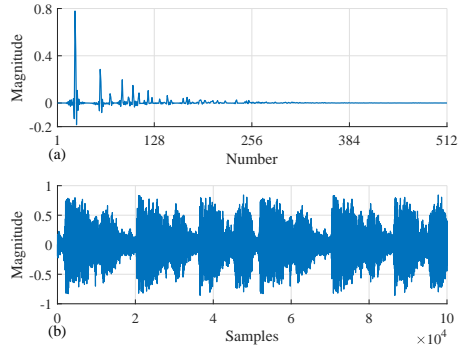}
\caption{(a) Room impulse response. (b) Speech input signal.}
\label{fig:acoustic_echo}
\end{figure}

\begin{figure}[!t]
\centering
\includegraphics[width=2.6in]{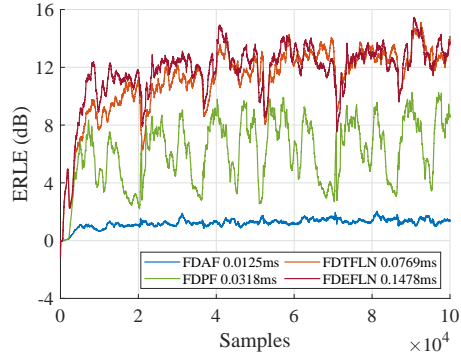}
\caption{Comparison of adaptive algorithms in the NAEC scenario, with $M = 64$, $\mu_w = 0.00005,\mu_q = 0.09$ for FDEFLN, $\mu = 0.00005$ for FDTFLN, and $\mu = 0.001$ for FDPF and FDAF.}
\label{fig:comparison_NAEC}
\end{figure}

We investigate a realistic NAEC scenario to check the performance of adaptive algorithms. The room impulse response of the acoustic echo path is modeled by the number of 512 taps \cite{Yu2016Novel}, and a speech input signal with an 8000Hz sampling rate is characterized by the magnitude with the range of $(-1, 1)$, whose magnitudes are illustrated in Fig. \ref{fig:acoustic_echo}. We have considered a memoryless sigmoidal nonlinearity to formulate an asymmetric loudspeaker distortion as $\phi(u(n)) = \beta/[1 + e^{-\alpha \gamma(n)}] - 0.5\beta$ where $\beta = 2$ is the gain, and $\alpha$ is the slope given by $\alpha= 4$ if $\gamma(n)>0$ and $\alpha= 0.5$ if $\gamma(n)<0$ with $\gamma(n) = 1.5u(n)-0.3u^2(n)$ \cite{Park2014Frequency}. We compare the performance of FDEFLN with that of the above-mentioned adaptive algorithms. The performance metric considered in the NAEC scenario is the echo return loss enhancement (ERLE) defined as $\mathrm{ERLE} = 10\log10(\mathbb{E}\{d^2(n)\}/\mathbb{E}\{e^2(n)\})$. The comparison of adaptive algorithms is shown in Fig. \ref{fig:comparison_NAEC}. Apparently, the linear FDAF algorithm is invalid. It is also founded that FDEFLN exhibits better ERLE level than other frequency domain algorithms in this real NAEC scenario.

\subsection{Implementations of FDEFsLMS}

\subsubsection{NANC with the engine noise}

We have investigated the implementation of NANC scenarios to examine the performance of FDEFsLMS. The reference signal $u(n)$ is produced by real-world engine noise data with an 8000Hz sampling frequency \cite{Lee2009Subband}. The primary noise is considered by a primary path nonlinearity at the cancellation point and is provided by $d(n) = \hat u(n-2) + 0.8 \hat u^2(n-2) - 0.4 \hat u^3(n-2) + \hat u^4(n-1) \hat u(n-2)$, where $\hat u(n) = u(n)*p(n)$ with $p(n)$ being the impulse response of $P(\mathrm{z}) = \mathrm{0.8z^{-9} + 0.6z^{-10}- 0.2z^{-11} - 0.5z^{-12} - }$ $\mathrm{ 0.1z^{-13} + 0.4z^{-14} - 0.05z^{-15}}$. The secondary path is assumed to be perfectly obtained through an offline modeling process, and it is considered to be $S(\mathrm{z}) = \hat S(\mathrm{z}) =\mathrm{ z^{-5} +2.5z^{-6} +1.76z^{-7} + 0.15z^{-8} - 0.4825z^{-9} - }$ $\mathrm{0.18625z^{-10}- 0.005z^{-11} - 0.001875z^{-12}}$. These experimental configurations have been commonly used in NANC literatures \cite{Le2018Generalized,Chen2021Active}.

\begin{figure}[!t]
\centering
\includegraphics[width=2.7in]{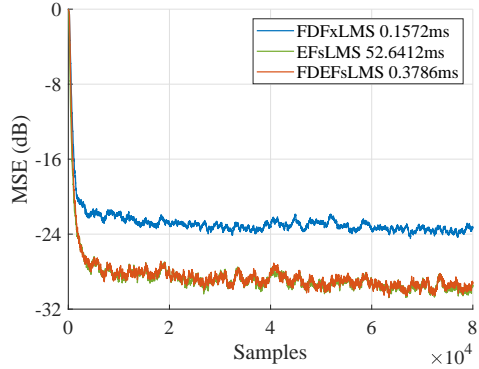}
\caption{MSE curves for the NANC system, with $M=100$, $\mu_w = 0.00001,\mu_q = 0.0001$ for FDEFsLMS and EFsLMS, and $\mu = 0.00001$ for FDFxLMS.}
\label{fig:comparison_ANC}
\end{figure}
\begin{figure}[!t]
\centering
\includegraphics[width=2.7in]{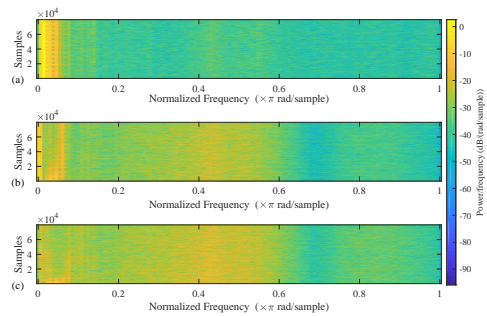}
\caption{Comparison of spectrograms. (a) Engine noise. (b) Residual error of FDFxLMS. (c) Residual error of FDEFsLMS.}
\label{fig:error_spectrogram}
\end{figure}
Fig. \ref{fig:comparison_ANC} illustrates the comparative MSE curves for the linear frequency domain filtered-x least mean-square (FDFxLMS) and EFsLMS-based algorithms. Fig. \ref{fig:error_spectrogram} shows a comparison of normalized frequency between the reference signal and the residual errors of FDFxLMS and FDEFsLMS. It is found that the convergence performance is almost the same in both the FDEFsLMS and EFsLMS algorithms, and they have better performance than the traditional linear FDFxLMS in the NANC scenario. The results of average execution time also show a significant improvement in computational efficiency of FDEFsLMS using 0.3786ms as compared to EFsLMS using 52.6412ms.

\subsubsection{NANC with the practical acoustic path}

\begin{figure}[!t]
\centering
\includegraphics[width=2.5in]{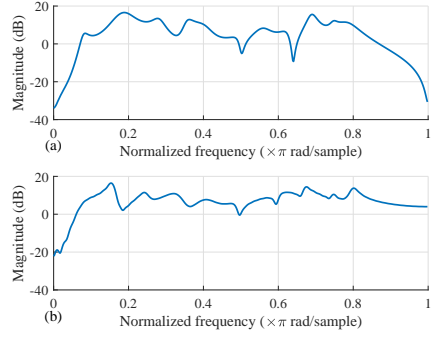}
\caption{Magnitude of frequency response. (a) Portion of primary path. (b) Secondary path.}
\label{fig:frequency_response}
\end{figure}

In this experiment, we consider a NANC scenario with the practical acoustic path, where the magnitude of frequency response for the portion of primary path $P(\mathrm{z})$ and the secondary path $S(\mathrm{z})$ is depicted in Fig. \ref{fig:frequency_response} \cite{Patel2016Compensating}. The whole nonlinear primary path is a cascade of a linear model $P(\mathrm{z})$ and a nonlinear microphone distortion provided by $d(n) = \beta/[1 + e^{-\alpha \hat u(n)}] - 0.5\beta$ with $\alpha = 6$ and $\beta = 7$. The reference signal $u(n)$ is a sinusoidal wave produced by $u(n) = \sin (2 \pi f n/f_s) + v(n)$ with $f=500$Hz and $f_s=4000$Hz and a $\mathrm{SNR} = 40$dB Gaussian noise $v(n)$ \cite{Patel2016Design}.

\begin{figure}[!t]
\centering
\includegraphics[width=2.5in]{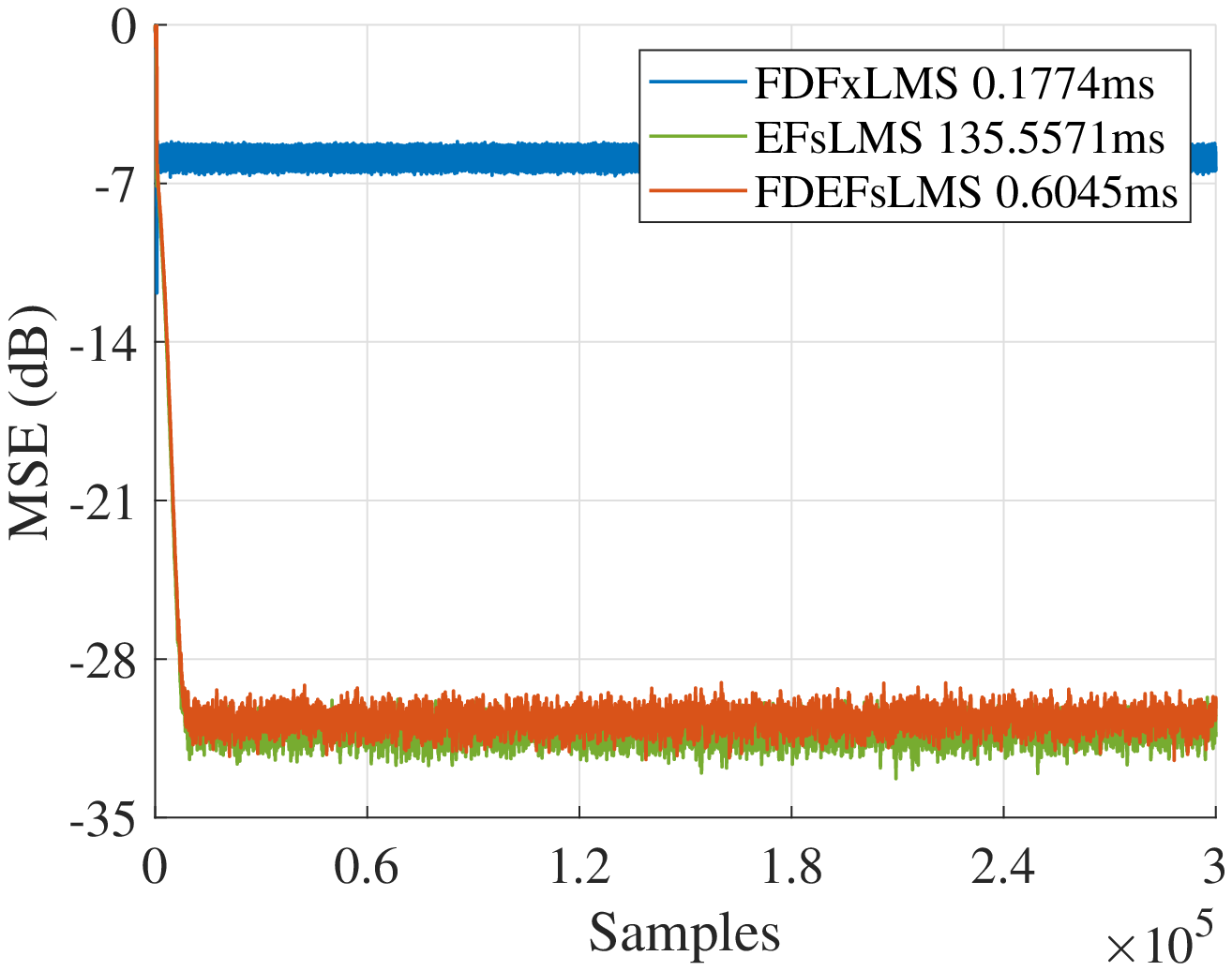}
\caption{MSE curves for the NANC system, with $M=100$, $\mu_w = 0.00002,\mu_q = 0.002$ for FDEFsLMS and EFsLMS, and $\mu = 0.00002$ for FDFxLMS.}
\label{fig:comparison_ANC_real}
\end{figure}
\begin{figure}[!t]
\centering
\includegraphics[width=2.5in]{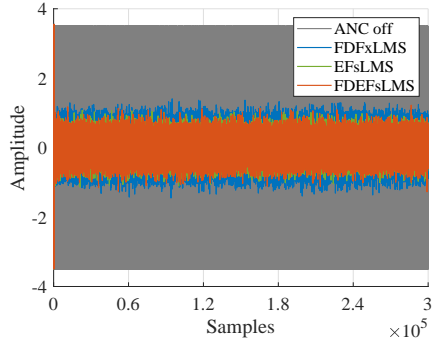}
\caption{Denoising results of adaptive algorithms.}
\label{fig:denoising_result}
\end{figure}
Fig. \ref{fig:comparison_ANC_real} shows the comparative MSE curves of frequency domain algorithms, and the denoising results with residual errors can be seen from Fig. \ref{fig:denoising_result}. We can obviously see that the linear FDFxLMS algorithm has a severe performance degradation, and the convergence properties of FDEFsLMS and EFsLMS are almost uniform. It means that the frequency domain implementation does not change the convergence behavior compared to the corresponding time domain algorithm. However, the proposed FDEFsLMS algorithm consumes 0.6045ms for each data block, and EFsLMS consumes 135.5571ms for $M$ samples, which further verifies that FDEFsLMS can evidently decrease the computational requirement on the premise of good convergence property.

\subsubsection{NANC with the chaotic noise}

In order to examine the tracking ability of the proposed algorithm, we investigate a NANC system with a nonlinear noise sequence and a nonminimum-phase response of secondary path. Since the logistic chaotic signal is a predictable and deterministic nonlinear process, the reference signal is often selected to be a logistic chaotic noise produced by $u(n+1) = \kappa u(n)[1- u(n)]$ with $\kappa=4$ and $u(0)=0.9$. This nonlinear reference noise is then normalized to have unit signal power \cite{Das2004Active}. In this scenario, the primary path model is considered to be $P(\mathrm{z}) = \mathrm{z^{-5} + 0.3z^{-6} + 0.2z^{-7}}$, and the portion of secondary path is taken as a nonminimum-phase model $S(\mathrm{z}) = \hat S(\mathrm{z}) = \mathrm{z^{-2} + 1.5z^{-3} - z^{-4} }$. Besides, the secondary path is prone to nonlinear distortion exhibited by $\hat y(n) = 3.3\tanh( 0.3y(n)*s(n) )$ \cite{Zhou2007Efficient,Patel2015Nonlinear}.

\begin{figure}[!t]
\centering
\includegraphics[width=2.7in]{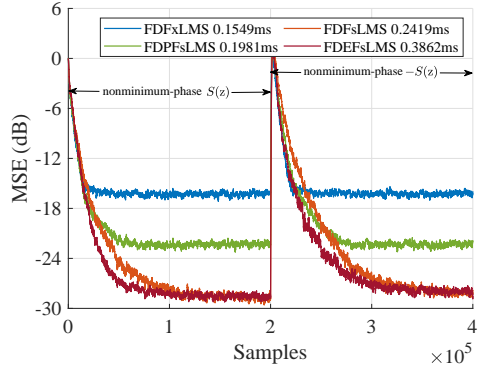}
\caption{Tracking abilities of adaptive algorithms, with $M=100$, $\mu_w = 0.00002,\mu_q = 0.0004$ for FDEFsLMS, $\mu = 0.00002$ for FDFsLMS, $\mu = 0.000025$ for FDPFsLMS, and $\mu = 0.00004$ for FDFxLMS.}
\label{fig:comparison_ANC_tracking}
\end{figure}

\begin{figure}[!t]
\centering
\includegraphics[width=2.7in]{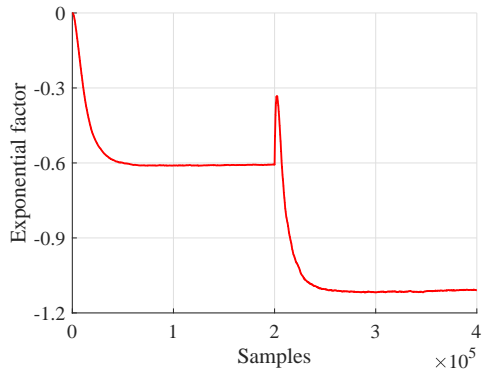}
\caption{Variations of the exponential factor for FDEFsLMS.}
\label{fig:q_result}
\end{figure}

To further assess the tracking capabilities of adaptive algorithms, the secondary path model has been suddenly changed to be opposite in sign at the half samples. As a fair comparison, the adaptive algorithms considered are the filtered versions with FDAF, FDPF, FDTFLN and FDEFLN for NSI, i.e., corresponding FDFxLMS, FDPFsLMS, FDFsLMS and FDEFsLMS for NANC. The MSE evolutions of adaptive algorithms with an abrupt secondary path model are shown in Fig. \ref{fig:comparison_ANC_tracking}. Fig. \ref{fig:q_result} shows the convergence result of the exponential factor for the FDEFsLMS algorithm. As can been seen, nonlinear adaptive algorithms have prominent performance advantages over linear FDFxLMS algorithm. We also find that the remarkable advantage of FDEFsLMS in terms of the convergence behavior, especially quicker tracking on the premise of the identical steady-state MSE, compared to the traditional FDFsLMS algorithm.

\section{Conclusion}

We have proposed a novel FDEFLN-based nonlinear filtering scheme and its filtered-s version, called the FDEFsLMS algorithm, which utilizes the block and FFT strategies in a computationally efficient manner. The bounds on step sizes, the details of the mean-square performance analysis, and the computational cost of the proposed frequency domain algorithm are given. With applications to NSI, NAEC and NANC, the notable computational advantages of the proposed FDEFLN-based algorithms have been confirmed, which all demonstrate the significant reduction of computational complexity in comparison with corresponding time domain algorithms. Moreover, the performance superiority of the proposed algorithms has also been verified over some existing algorithms.

\section*{Acknowledgments}

This work was supported in part by the National Natural Science Foundation of China under Grant 61901400, the Young Scholars Development Fund of SWPU under Grant 201899010157, and the Scientific Research Starting Project of SWPU under Grant 2019QHZ015.

\section*{References}
\bibliographystyle{elsarticle-num}
\setlength{\bibsep}{0.5ex}
%\biboptions{sort}
\bibliography{SPref}

\end{document}